\documentstyle[12pt]{article}
\def\ot#1#2{\textstyle \frac{#1}{#2}}
\def\od#1#2{\displaystyle \frac{#1}{#2}}
\def\beq{\begin{equation}}
\def\eeq#1{\label{#1}\end{equation}}
\def\barr#1{\begin{equation}\begin{array}{#1}\displaystyle}
\def\earr#1{\end{array}\label{#1}\end{equation}}
\def\dis{&\displaystyle}
\def\di#1{\\[0.#1cm]\displaystyle}
\def\r#1{(\ref{#1})}
\def\si{\sigma}
\def\ga{\gamma}
\def\ep{\epsilon}
\def\al{\alpha}
\def\be{\beta}
\def\de{\delta}
\def\De{\Delta}

\def\la{\lambda}
\def\ta{\tau}
\begin{document}
\begin{center}
 { \Large \bf      POLRAD 2.0.  FORTRAN code  for the Radiative \\
  Corrections Calculation to Deep Inelastic \\
      Scattering  of  Polarized Particles \\[0.7cm]
 \large
I. Akushevich, A. Ilyichev, N. Shumeiko, A. Soroko, A.
Tolkachev \\[0.7cm]
 \it National Scientific and Education Center of Particle and \\
  High Energy Physics attached to Byelorussian State University
\\ Bogdanovich str. 153, 220040 Minsk, Belarus
\\[0.7cm]  \normalsize {\bf Corresponding author:}
I.Akushevich,
e-mail: {\it aku@hep.by}, phone/fax: 375 17 2326075
}
\end{center}

\vspace{0.5cm}

\begin {abstract}
The FORTRAN code POLRAD 2.0 for radiative correction
calculation in inclusive and semi-inclusive  deep
inelastic
 scattering of polarized leptons by polarized nucleons and
nuclei is described.
Its theoretical basis,
structure and algorithms are discussed in details.

\end{abstract}

\vspace{0.5cm}
\vspace{0.5cm}


\vspace{0.5cm}

\noindent {\bf Program Library Index:} Particle Physics Quantum
Electrodynamics

\newpage
\noindent
\section*{Program Summary}
{\it Title of program:} POLRAD
\newline
{\it Version:} 2.0 April 1997
\newline
{\it Catalogue identifier:}
\newline
{\it  Program  obtainable  from:}
on request from e-mail: aku@hep.by
\newline
{\it Computer for which the program is designed and others on which
it has been tested:}
\newline
{\it Computers:} all
\newline
{\it Operating systems or monitors under which the program has been
tested:} all
\newline
{\it Programming language used:} FORTRAN 77
\newline
{\it Memory required to execute with typical data:} 1MB
\newline
{\it No. of bits in a word:} 32
\newline
{\it No. of processors used:} 1
\newline
{\it No. of bytes in distributed program, including test data,
etc.:} 300 kB
\newline
{\it Distribution format:} default ASCII else uuencoded compressed
tar file
\newline
{\it Other programms called:}
\newline
PATCHY [1] --- part of CERNLIB
\newline
MINUIT [2] --- part of CERNLIB
\newline
{\it Keywords:} polarized particles, inclusive and
semi-inclusive deep inelastic scattering,
QED and electroweak radiative corrections, structure functions,
higher order corrections, experimental data processing.
\newline
{\it Nature of physical problem:}
\newline
First and higher order QED and electroweak radiative corrections
to the inclusive and semi-inclusive polarized deep inelastic
scattering; experimental data processing.
\newline
{\it Method of solution:} Numerical integration of analytical
formulae.
\newline
{\it Restrictions on complexity of the problem:}
\newline
Only selective experimental cuts are possible. For ${\cal
O} (\alpha^2)$ order correction only leading contribution is
calculated. Electroweak correction is calculated for longitudinally
polarized target.
\newline
{\it Typical running time:}
\newline
The running time depends on the options
used. For example: 1) calculation of the total QED correction
takes about 4 seconds of the CPU time per one kinematical point;
2) calculation of electroweak+
${\cal O} (\alpha^2)$+model for $g_2 \neq 0$ takes up to 300
seconds per one kinematical point.
\newline
{\it References:}
\newline
[1] H.J.Klein, J.Zoll, PATCHY Reference Manual, March 1988.
\newline
[2] F.James, MINUIT Reference Manual, March 1994.
\newline
\newpage

\section{Introduction}
Data processing of the modern experiments
on deep inelastic scattering (DIS) of polarized leptons on
polarized nuclear target
requires correct account of the radiative corrections (RC). Our
program POLRAD
2.0 based theoretically on the original approach proposed in
ref.\cite{KSh} and developed in the ref.\cite{ASh}
 was created to suit the demands of the present
 and future experiments with
fixed polarized nuclear targets and at collider.
Along with the possibilities of the previous versions of POLRAD
\cite {AShT}, which calculated the QED lowest order RC to DIS of
polarized  leptons by polarized nuclei, the current version gives
an opportunity to take into account both electroweak and higher
order effects and to calculate the RC for semi-inclusive polarized
experiments.

In section \ref{2.0} we present the detailed description of
the theoretical basis of POLRAD 2.0 along with the explicit
formulae.
We start with the calculation of born cross section in subsection
\ref{2.01}.
In subsections \ref{2.02} and \ref{2.2} we present the review of
the basic formulae for each of the
radiative tails: elastic, quasielastic and inelastic, ---
and consider the case of  ultrarelativistic
approximation to the lowest order QED correction.
The
contribution of $\alpha^2$ order correction is calculated in
subsection \ref{2.3}
on
the basis of structure functions formalism.
The expressions for one-loop electroweak correction within the
framework of standard theory and QCD-improved parton model
are given in subsection
\ref{2.4}.
The POLRAD 2.0 part that calculates RC in
semi-inclusive case is the modification of the
code SIRAD \cite {SSh} and is described in subsection \ref{2.5}.
The current data processing iteration procedure discussed in
subsection \ref{2.6}
is extended
in comparison with the one in the previous
version of POLRAD to fit the data using the CERNLIB
package MINUIT.

\ref{appA} is devoted to structure function definition,
parameterization and models used. Most cumbersome parts of explicit
formulae are presented in Appendices B and C.

The current version of POLRAD gives the opportunity to choose one of
five nuclear targets along with the type of its polarization, type of
scattered longitudinally polarized charged leptons. It is possible to
operate in standard SMC \cite{SMC}, HERMES \cite{HERMES}, E142
\cite{E142} kinematics or to choose any
other kinematics. Procedure of RC of experimental data can be organized
with implementation of iteration procedure.
The program is realized on Standard FORTRAN 77 and does not require
any changes when used under different computer platforms and
operation systems.

\section{Theory}  \label{2.0}

We consider the process of DIS of longitudinally polarized
charged leptons on longitudinally and transversely polarized
nuclear target
\begin{equation}
l+N  \longrightarrow  l'+ X,
\end{equation}
and semi-inclusive DIS (SIDIS) process when a hadron is measured
in coincidence with the scattered lepton.
The physical interpretation of the experimental data
requires the separation of the Born cross section
from background contributions known as radiative corrections,
which
originate from loop diagrams and
from processes with the emission of additional real photons.
 Radiative
events cannot completely be removed by experimental methods and so
they have to be calculated theoretically and substracted
from measured cross sections.

    It is well-known that there are three scattering channels of virtual
boson ($\gamma$, $Z$) on nucleus in  dependence on transfer energy
$\nu
=E_{1}-E_{2}$,
where $E_{1}(E_{2})$  is initial  (scattered) lepton  energy:   elastic,
quasielastic and inelastic.   Representative plot  of dependence of  the
scattering  cross  section  on  $\nu  $  and square of transfer momentum
$Q^{2}=-q^{2}$ is shown on fig.1 (only the regions that give
sufficient contribution to RC calculation). Peak in the range I
(for
$\nu =  0$,
if to  neglect nuclear  recoil) corresponds  to elastic  scattering.  In
this case the nucleus  remains in the ground  state.  Range II
stands for the
quasielastic scattering i.e. direct collisions of leptons with  nucleons
inside nucleus.  Wide maximum in the energy spectrum originates from the
own movement of nucleons.  Range III of inelastic scattering occurs when
transfer energy is greater then pion threshold.

On the born level
    both $\nu $ and $Q^{2}$ are  fixed by the measurements of the
scattered photon momentum.
Hence, the  channel of  scattering is  fixed too.
However,
 on  a  level  of  RC
radiated real photon momentum is indefinite, hence,
$\nu  $  and  $Q^{2}$ are arbitrary so
each of three channels
contributes to cross section.   Integration over the photon phase  space
may be presented to  that in  a plane  of $\nu  $ and
$Q^{2}$. Adding the virtual
photon contribution $\sigma ^{v}$, we have for the RC cross
section

\begin {equation}
\sigma  = \sigma ^{in}+ \sigma ^{el}+ \sigma ^{q}+ \sigma ^{v}.
\label{eq1}
\end {equation}

    Here each $\sigma  $ denotes the  double differential cross  section
$d^2\sigma / dxdy$, and  $x,y$ are  usual scaling  nucleon
variables.
$\sigma  ^{in}$,
$\sigma  ^{el}$, $\sigma  ^{q}$
are contributions  of radiative tails from continuous
spectrum  (IRT), of the elastic scattering radiative tail (ERT),
of the radiative tail from the quasielastic  scattering
(QRT) respectively.
Also the contribution of electroweak correction
calculated in the quark-parton model
is contained in
$\sigma  ^{in}$.
Both $\alpha$ and $\alpha^2$ corrections are taken into account in
$\sigma^{in,el,q,v}$. To separate the contributions we introduce
the lower index, f.e. $\sigma^{in}=\sigma^{in}_1+\sigma^{in}_2 $.

All above mentioned contributions are valid in the case of
inclusive scattering. However, in semi-inclusive case the
transfer energy
is above pion threshold, so in RC calculation
one have to take into account only $\sigma^{in}$ and
$\sigma^{v}$.

\subsection{QED formulae for inclusive case} \label{2.1}
\subsubsection{Born contribution}  \label{2.01}

Using (\ref{ht}) for the DIS cross section on the  Born  level, we
obtain
\begin {equation}
\begin {array}{l}
\displaystyle
{d\sigma \over {dxdy}} = {4\pi \alpha ^{2}\over {\lambda _{s}}}
 {SS_{x}\over Q^{4}}
 \left\{
  (Q^2-2m^{2})\Im _{1}+(SX-M^{2}Q^2){\Im _{2}\over
2M^2} \right. \\[0.5cm]
\displaystyle
\;\;\;\;\;\;\; + mMP_{L}
 \left(2(Q^2\;\xi \eta -q\eta \;k_{2}\xi ){\Im _{3}\over
  M^2}+ (S_{x}\;k_{2}\xi -2\;\xi p\;Q^2)
 q\eta {\Im _{4}\over M^4} \right) \\[0.5cm]
\displaystyle
\;\;\;\;\;\;\; +  (Q^2-2m^{2})(Q^2-3(q\eta )^2){\Im _{5}\over
 M^2}+ (SX-M^2Q^2)(Q^2-3(q\eta )^{2}){\Im _{6}\over
 2M^4}     \\[0.5cm]
\displaystyle
\;\;\;\;\;\;\; \left.
 - {1\over 2}\pmatrix{Q^2+4m^2+12\;\eta k_1\;\eta
 k_2}\Im _{7} - {3\over 2}(X\;\eta k_{1}+S\;\eta k_{2})
 q\eta{\Im _{8}\over M^2}
 \right\} .
\end {array}
\label{bt}
\end {equation}
Here $k_{1}(k_{2}),\xi ,m$  are  initial  (final)  lepton   momentum,
polarization vector and its mass respectively.
 Invariants are defined  in
a standard way

\begin{equation}
\begin{array}{c}
\displaystyle
S = 2k_{1}p,\; X = 2k_{2}p = (1-y)S,\; Q^2 = -(k_{1}-k_{2})^{2} ={ xyS},
\\[0.5cm]\displaystyle
S_{x}= S-X,\;Q^2_m=Q^2+2m^2,\;S_p=S+X,\; \lambda _{s}=
S^{2}-4m^{2}M^{2},
\end{array}
\label{invar}
\end{equation}
$P_L$ is initial lepton
polarization degree. An explicit form of hadronic tensor and
generalized structure functions $\Im_i$ are presented in Appendix
\ref{appA1}.

The equality (\ref{bt}) is true  for  the
any direction of polarization vectors and is exact:  no
approximations were made yet.

    The  4-vector  of  target  polarization  $\eta$  is  the   covariant
representation of target polarization vector $\vec{n}$.
In the lab. frame
$\eta
=
(\vec{n}$,0),
and $\vec{n}$  can be  expanded in  three  components:   parallel
to
initial lepton momentum
$\vec{k}_{1}$  -  $\vec{n}_{L}$,  normal  to $\vec{k}_{1}$ in scattering
plane  $(\vec{k}_{1},\vec{k}_{2})$   -  $\vec{n}_{t}$   and  normal   to
scattering plane -  $\vec{n}_{\bot }$.   If a target  is polarized along
$\vec{n}_{L}(\vec{n}_{t})$,  then  we   speak  about  the   longitudinal
(transverse)  polarization.    In  the  third  case  we  speak about the
polarization being normal to scattering plane.

    Let us build  a basis in  the 4-dimensional space.   The process  of
inclusive  DIS  is  determined  by  three vectors of incoming (outgoing)
lepton  $k_{1}(k_{2})$  and  of  incoming  nuclei  $p$  defining a
hyperplane in the  4-dimensional space.   We can choice  the orthonormal
basis system in this hyperplane  $( {p\over M}, \eta _{L},  \eta _{T})$,
where

\begin{equation}
\begin{array}{c}
\displaystyle
\eta _{L}= \lambda ^{-1/2}_{s}\pmatrix{2Mk_{1}-{S\over M}p},
\\[0.5cm]\displaystyle
\eta _{T}= {(-SX+2M^2Q^2_m))k_{1} + \lambda _{s}k_{2}
-(SQ^2+2m^{2}S_{x}) p\over
\lambda ^{1/2}_{s} (SXQ^2 - m^{2}S^{2}_{x}-M^{2}Q^4-4m^2M^2Q^2)^{1/2}}.
\end{array}
\end{equation}

    Among all possible basis system,  our system differs from others  in
following:  two space-like vectors  $\eta _{L}$ and $\eta _{t}$  in lab.
frame have the form  $(\vec{n}_{L,t},0)$,  where  $\vec{n}_{L,t}$
are
above-considered 3-dimensional polarization  vectors.  The  basis vector
system  is  uniquely   fixed  by  this   requirement.    Basis   in  the
4-dimensional space is produced by adding to the system a
 4-momentum  $\eta _{\bot }  ((\vec{n}_{\bot },0)$ in  the
lab system)
orthonormal to
the hyperplane.

As a result for any 4-vector $\eta $ we have expansion
\begin{equation}
\eta = {\eta p\over M} {p\over M} - (\eta \eta _{L})\eta _{L}- (\eta
\eta _{t})\eta _{t}- (\eta \eta _{\bot })\eta _{\bot } .
\label{A.4}
\end{equation}

    If  $\eta  $   is    target polarization  vector,  then  for   three
above-mentioned  cases  we  find:    $\eta  =\eta  _{L}$   (longitudinal
polarized target),  $ \eta  =\eta _{t}$  (transversely polarized
target),
$\eta =\eta  _{\bot }$  (target polarized  orthogonal to  the scattering
plane).

    Initial lepton is  always longitudinally polarized  (for
experiments
considered).  Using the expansion (\ref{A.4}) for this vector we obtain

\begin{equation}
\xi  = \xi _{L}= \lambda ^{-1/2}_{s} \pmatrix{{S\over
 m}k_{1}-2mp}.
\end{equation}

    We note here, that calculation of real photon contribution  requires
to  integrate  over  $d^{3}\vec{k}$  ($k$  -  real photon momentum) some
expressions,  containing  the  scalar  products  of $k$ and polarization
vectors $\xi  $ and  $\eta $.  Since scalar  products $k\xi , k\eta_{L},
k\eta  _{t}$  are  easily  expressed  in  terms  of invariants, then our
treatment allows  to eliminate  the intricate  and tedious  procedure of
tensor integration used in ref.\cite {KSh} and significantly  simplifies
results.    This  is  the  most  important  advantage  of the considered
treatment.

By applying  the  ultrarelativistic  approximation
\begin {equation}
m^2,M^2 \ll S,X,Q^2
\end {equation}
  and
making transfer to scaling variables $x$ and $y$ we further find
\begin {equation}
\begin {array}{c}
\displaystyle
{d\sigma \over {dxdy}}={4\pi \alpha ^{2}S\over Q^4}
((F_{1}-{Q_{N}\over 3}b_{1})xy^{2}+(F_{2}-{Q_{N}\over 3}b
 _{2})(1-y)  \\[0.5cm]
\displaystyle
-P_{L}P_{N}xy(2-y)g_{1})
\end {array}
\label {bl}
\end {equation}
in the case of longitudinal polarized target and
\begin {equation}
\begin {array}{c}
\displaystyle
{{d\sigma \over {dxdy}}} = {{4\pi \alpha ^{2}S\over Q^4}}
((F_{1}+{Q_{N}\over 6}b_{1}
 )xy^{2} + (F_{2}+{Q_{N}\over 6}b
 _{2})(1-y)          \\[0.5cm]
\displaystyle
-2P_{L}P_{N}{x\sqrt {xy(1-y)}M\over \sqrt {S}}
(yg_{1}+2g_{2}))
\end {array}
\label {br}
\end {equation}
for transverse one.

\subsubsection{Exact formulae of the lowest order} \label{2.02}
The model independent RC of the lowest order can be written as
the sum of bremsstrah\-lung and loop effects:
\begin {equation}
\sigma  = \sigma ^{in}_1+ \sigma ^{el}_1+ \sigma ^{q}_1+ \sigma
^{v}_1.
\label{eq1a}
\end {equation}

The explicit form for these contributions was obtained in
ref.\cite{ASh}. For the infrared free sum of $\sigma ^{v}_1$
 and $\sigma ^{in}_1$ we have
\begin {equation}
\sigma ^{v}_1 + \sigma ^{in}_1= {\alpha \over \pi }
\delta_{v}
 \sigma _{o} + \sigma ^{in}_F.
= {\alpha \over \pi }
(\delta^{IR}_{R} + \delta _{vert} + \delta _{vac}^l + \delta _{vac}^h)
 \sigma _{o} + \sigma ^{in}_F.
\label{RpV}
\end {equation}
$\sigma ^{in}_F$ is the infrared free part of the IRT cross section
\begin {equation}
\begin {array}{l}
\displaystyle
\sigma ^{in}_F = -{\alpha ^{3}y}
 \int\limits^{\tau_{max}}_{\tau_{min}}d\tau \sum^{8}_{i=1}
 \left\{ \theta _{i1}(\tau)
\int\limits^{R_{max}}_{0}{dR\over R} \right.
\left[ {\Im _{i}(R,\tau )\over (Q^2+R\ta)^2}-{\Im _{i}(0,0)\over
Q^4}\right]\\[0.5cm]
\displaystyle
\;\;\;\;\; \qquad \qquad\qquad\qquad\qquad\left.
+ \sum^{k_{i}}_{j=2} \theta _{ij}(\tau)\int\limits^{R_{max}}_{0}
dR {R^{j-2}\over (Q^2+R\ta)^{2}}\Im _{i}(R,\tau)  \right\}.
\end {array}
\label{eq25}
\end {equation}
The integration region on
variables $R=2pk$ and $\tau=k(k_1-k_2)/pk$ is sketched on
fig.\ref{limint}a. The limits of integration are defined as
\begin {equation}
\begin {array}{c}
 \displaystyle
R_{max} ={{W^{2}-(M+m_{\pi })^2}\over 1+\tau},  \ \
 \displaystyle
\tau _{max,min} = {S_{x} \pm \sqrt {\lambda _{Q}}\over 2M^{2}},
\\[0.5cm] \displaystyle
\lambda _{Q}= S^{2}_{x}+ 4M^{2}Q^2,\ \  W^{2}= S_{x}- Q^2 +
M^{2},
\end {array}
\end {equation}
where $m_{\pi}$ is the pion mass.
The explicit form of functions $\theta _{ij}(\ta )$ is given in
\ref{appB}.

The quantity $\delta ^{IR}_{R}$  appears  when the infrared
divergence
is extracted in accordance with  the Bardin and
Shumeiko
method \cite{BSh}
from $\sigma ^{in}$.
The  virtual  photon  contribution consists of the
lepton
vertex correction $\delta _{vert}$  and  the  vacuum  polarization  by
leptons $\delta ^{l}_{vac}$  and  by  hadrons $\delta ^{h}_{vac}$
\cite{delvac}.
These corrections are given by formulae (20-25) of ref.\cite{ASh}.
Here we give ultrarelativestic formulae ($m\rightarrow 0$) for sum
of
$\delta ^{IR}_{R}$ and  $\delta _{vert}$:
\begin {equation}
\delta ^{IR}_{R} + \delta _{vert}=\delta_{inf}
+\od32\l_m-2-\od12\ln^2\od XS+{\rm Li}_2
\od{SX-Q^2M^2}{S'X'}
-\od{\pi^2}6,
\end {equation}
where $S'=X+Q^2$, $X'=S-Q^2$, $l_m=\ln Q^2/m^2$ and ${\rm Li}_2$
is Spence function (dilogarithm) and
\begin {equation}
\delta _{inf}=
(l_m-1)\ln
\od{(W^2-(M+m_{\pi})^2)^2}{S'X'}.
\end {equation}

In the case of elastic scattering the nucleus remains in the ground
state, so we have an additional relation
\beq
R=R_{el}=(S_{xA}-Q^2)/(1+\ta_A)
\eeq{Rel}
resulting in
\begin {equation}
\sigma_1^{el}= {1\over A}{d^2\sigma ^{el}\over dx_Ady}=
- {\alpha ^{3}y\over A^2}\int\limits^{\tau_{Amax}}_{\tau_{Amin}}d\tau_A
 \sum^{8}_{i=1}\sum^{k_{i}}_{j=1} \theta _{ij}(\tau_A){
 2M^{2}_A R^{j-2}_{el}\over (1+\tau_A)(Q^2+R_{el}\ta_A)^{2}}\Im
^{el}_{i}(R_{el},\tau_A).
\label{eq23}
\end {equation}
Here invariants with the index "A"
 contain the nucleus momentum
$p_A$ instead of $p$ ($p_A^2=M_A^2$, $M_A$ is nucleus mass).
The quantities $\Im ^{el}_i$ are given in Appendix \ref{appA1}.

Quasielastic scattering corresponds to direct collisions of leptons
with nucleons
inside nucleus.  Due to
self movement of nucleons we have no additional relation like \r{Rel}.
As a result we have to integrate numerically both over $R$ and $\ta$
\begin {equation}
\sigma ^{q}_1= - {\alpha ^{3}y\over
A}\int\limits^{\tau_{max}}_{\tau_{min}} d\tau\sum^{8}_{i=1}
 \sum^{k_{i}}_{j=1} \theta _{ij}(\tau)\int\limits^{R_{max}^q}
 _{R_{min}^q} dR
 {R^{j-2}\over (Q^2+R\ta)^2}\Im_{i}^q(R,\tau).
\label {qu}
\end {equation}
The quantities $\Im ^q_i$ can be obtained in the terms of quasielastic
structure functions (so-called response functions, see Appendix
\ref{appA1} for explicit result), which have a form of the peak
for
$\omega
=Q^2/2M$. Due to the absence of enough experimental data the
fact is normally used for construction of the peak type approximation.
The factors at
response functions
are estimated at the
peak, and
subsequent integration of response functions leads to results in terms
of suppression
factors $S_{E,M,EM}$ (or of sum rules for electron-nucleus scattering
\cite{LLS}):
\begin {equation}
\sigma ^{q}_1=
- {\alpha ^{3}y\over A}\int\limits^{\tau_{max}}_{\tau_{min}}d\tau
 \sum^{4}_{i=1}\sum^{k_{i}}_{j=1} \theta _{ij}(\tau){
 2M^{2} R^{j-2}_{q}\over (1+\tau)(Q^2+R\ta)^{2}}\Im
^{q}_{i}(R_{q},\tau).
\label{eq23q}
\end {equation}

To take into account the effect of radiation of many soft photons
a special procedure of exponentiation was applied \cite{Sh}. In
the code it is realized by the following substitutions:
\barr{c}
 \si^{el}_1\rightarrow \left({y^2(1-x/A)^2\over
1-xy/A}\right)^{t_r}\si^{el}_1,
\qquad
 \si^{q}_1\rightarrow \left({y^2(1-x)^2\over
1-xy}\right)^{t_r}\si^{q}_1,
\di5
 (1+\od{\al}{\pi}\delta_v)\si_o \rightarrow
 \exp\left(\od{\al}{\pi}\de_{inf}
 \right)\bigl(1+\od{\al}{\pi}(\delta_v-\delta_{inf})\bigr)\si_o,
\earr{00721}
where $t_r= \od{\al}{\pi}(l_m-1)$.

\subsubsection{Ultrarelativistic approximation}        \label{2.2}
To simplify and accelerate the procedure of experimental data
processing when rapid ana\-ly\-sis is more important than
accuracy, it is convenient to have the approximate formulae.
In the case of
RC calculation one can choose
ultrarelativistic
approximation:
\begin{equation}
m^2, M^2 \ll S,X,Q^2,
\label{apr}
\end{equation}
that allows to calculate exactly first two terms (corrections
$\sim \alpha l_m$ and
$\sim \alpha$) of expansion over the leptonic mass $m$ of the
lowest order cross section.

The expressions derived under such an approximation are compact
and have good accuracy.
Also this
approach allows to avoid numerical uncertainties when the value
results from
the difference between two large and sometimes infinite
quantities,
that is especially  significant when quadruple polarization is
considered.

\subsubsection*{A. Inelastic radiative tail}
Considering $\tau$-dependence of quantities $\theta_{ij}(\tau)$
in (\ref{eq25}) one
can see its peaking structure, so called $s$- and $p$-peaks
\cite{MT} (or $k_1-$ and $k_2-$peaks if follow \cite{Sh}):
$\theta_{ij}(\tau)\sim \theta_{ij}^s(\tau)+\theta_{ij}^p(\tau)$.
Using the identities
\begin{equation}
\begin{array}{c}
\theta_{ij}^s(\tau)\Im_i(R,\tau)=\theta_{ij}^s(\tau)(\Im_i(R,\tau)-
\Im_i(R,\tau_s))+\theta_{ij}^s(\tau)\Im_i(R,\tau_s), \\ [0.5cm]
\theta_{ij}^p(\tau)\Im_i(R,\tau)=\theta_{ij}^p(\tau)(\Im_i(R,\tau)-
\Im_i(R,\tau_p))+\theta_{ij}^p(\tau)\Im_i(R,\tau_p),
\label{theta}
\end{array}
\end{equation}
one can extract and analytically integrate the terms corresponding
the mass singularities. The first terms in
right-hand sides of
(\ref{theta}) are free from mass singularities and so
they contribute only to $\sim\alpha$ correction. So one can adopt
$m^2=M^2=0$
before the integration over $\tau$ (or photon radiation angles).
SF's do not depend on $\tau$ in the rest terms.
Hence, the last ones
can be integrated
analytically using the methods \cite{BSh2}
and contribute to the leading correction $\sim \alpha l_m$.

Results for the infrared free sum of contributions from inelastic
radiative tail and loop effects could be presented as following
\beq
\sigma ^{in}_1+\sigma^v_1= \od {\alpha }{\pi }\delta^{in}_1\sigma_0
+\sigma^V_r
+\sigma _s^{in} +\sigma _p^{in}+\sigma _r^{in}.
\eeq{0004}
Factorizing terms of the total correction have the form
\barr{c}
\delta^{in}_1=
\od 1 2[(l_m-1)\delta_{sp}-\ln^2(1-y)-1]
,
\earr{0020a}
where
\barr{c}
\delta_{sp}=2\ln((1-z_p)(1-z_s))+3,
\earr{002a}
The quantity $\sigma^V_r$ is correction due to vacuum polarization
effects by leptons and hadrons:
\beq
\sigma^V_r
=(\delta^l_{vac}+\delta^h_{vac})\sigma_0
.
\eeq{0020b}

Quantities $\sigma _{s,p}^{in}$ are the contributions including
the second terms in right-hand side of (\ref{theta}):
\barr{l}
\sigma _{s,p}^{in}=\frac{\alpha}{2\pi }
\int\limits_{z_{s,p}}^{1}
\frac{dz}{1-z}
\left(
[(1+z^2)l_m-2z]\sigma _{s,p}
-2 (l_m-1)\sigma_0
\right)
,
\earr{sppeak}
The quantities $\si_{s,p}$ are expressed in terms of born
cross section
$\sigma _0=\sigma _0(S,X,Q^2)$:
\barr{c}
\sigma _s=y\sigma _0(zS,X,zQ^2)/(z-1+y) , \quad
\sigma _p=y\sigma _0(S,X/z,Q^2/z)/z(z-1+y).
\earr{0003}

The low limits of integration in \r{sppeak} are
\beq
z_s={1- y \over 1-xy}, \quad
z_p={1- y+xy}
.
\eeq{uplim}

The mass singularity free terms in
\r{theta} contribute to $\sigma _{r}^{in}$.
After splitting the unpolarized, polarized and
quadrupolarized parts, we have
\beq
\sigma _{r}^{in} = 2\alpha^3y\int\limits_x^1
\frac{d\xi}{\xi^2}
\left(
T_u^L+
P_LP_N (T_{p\| }^L+T_{p\| }^x)+
{\textstyle \frac{Q_N}{3}} T_q^L
\right)
\eeq{0009a}
in the case of longitudinally polarized target and
\beq
\sigma _{r}^{in} = 2\alpha^3y\int\limits_x^1 \frac{d\xi}{\xi^2}
\left(
T_u^L+
P_LP_N (T_{p\bot }^L+T_{p\bot }^x)-
{\textstyle \frac{Q_N}{6}} T_q^L
\right)
\eeq{0009b}
for transversely polarized target.
Quantities $T$ are

\def\DLYY{L^Y}
\def\DLTT{L_t}
\def\DLSK{L^k_s}
\def\DLXK{L^k_x}
\def\DLSL{L^l_s}
\def\DLXL{L^l_x}
\def\x{x_\xi }

\barr{ll}
T^L_u=  {\cal L}_1( F_1)+{1\over \xi Q^4_\xi }  {\cal L}_2(F_2)
,\dis
T^L_q=  {\cal L}_1(b_1) + {1\over \xi Q^4_\xi }  {\cal L}_2(b_2)
,\di5
T^L_{p\| }=-\od{1}{Q^2_{\xi }} {\cal L}_3(g_1)
,\;\;\;\;\;\;\;\;\;\;\;\;\;\;\;\;\;\;\dis
T^L_{p\bot }= \od{\xi M}{(SXQ^2)^{1/2}}({\cal L}_4(g_1)
+2{\cal L}_5(g_2)/Q^2_\xi)
,\di5
T^x_{p\| }= {2u(S+u_x)g_1(\xi,t_x)\over u_xXQ^2_\xi }
,\;\;\;\;\;\;\;\;\;\;\;\;\;\;\dis
T^x_{p\bot }= {4\xi Mu(Q^2_\xi g_1(\xi ,t_x)+2Sg_2(\xi ,t_x))\over
XQ^2_\xi (SXQ^2)^{1/2}}
,
\earr{TTT}
where
\barr{l}
{\cal L}_1({\cal F})=2\DLYY -\DLSK +\DLSL +\DLXK -\DLXL +2\DLTT
,\di5
{\cal L}_2({\cal F})=
 (T-Q^2_\xi S_x)\DLYY-(Q^2_\xi X + T)\DLSK+(T-Q^2_\xi S)\DLXK
,\di5
{\cal L}_3({\cal F})=2S_p\DLYY -(u_s+X) \DLSK - Q^2_\xi  \DLSL
+ (S+u_x) \DLXK -Q^2_\xi  \DLXL
,\di5
{\cal L}_4({\cal F})=
-4X\DLYY +(2X-S)\DLSK -u_s\tilde \DLSK +(2Q^2_\xi-u_s  ) \DLSL
\di5 \qquad \qquad
-(X+2u_x)\DLXK- u_x\DLXL -u_x\tilde \DLXK+2S_x \DLTT
,\di5
{\cal L}_5({\cal F})=
(S_xQ^2_\xi-S_p^2   )\DLYY
+( SS_p +2Xu_s)\DLSK +u_sX\tilde \DLSK+
\di5 \qquad \qquad
- (2Su_x+ S_pX)\DLXK -u_xS\tilde \DLXK
.
\earr{call}
Here $u=S_x-Q^2_{\xi}$, $u_x=S-Q^2_{\xi}$, $u_s=X+Q^2_{\xi}$,
$Q^2_{\xi}=Q^2/\xi$, $T=S^2+X^2$.
Quantities $L$, defined below, have to be computed for the same
argument $\cal F$ (${\cal F}=F_{1,2}$, $g_{1,2}$, $b_{1-4}$) as
in
$\cal L(F)$.

\barr{l}
L_t=\od1{S_x}\int\limits_{t_1}^{t_2}dt
\frac{{\cal F}_i(\xi ,t)}{t}
,
\di5
L_{s,x}^l=\od{{\cal F}_i(\xi ,t_{s,x})}{u_{s,x}} \ln \od
{u_{s,x}^2}{uQ^2_{\xi}}+
\od1{u_{s,x}} \int\limits_{t_1}^{t_2}dt
\frac{{\cal F}_i(\xi ,t)-{\cal F}_i(\xi ,t_{s,x})}{|t-t_{s,x}|}
,
\di5
L_{s,x}^k=\od{{\cal F}_i(\xi ,t_{s,x})}{S,X} \ln \od
{u_{s,x}^2}{uQ^2_{\xi}}+\od1{S,X}\int\limits_{t_1}^{t_2}dt
\frac{t_{s,x}{\cal F}_i(\xi ,t)-t{\cal F}_i(\xi ,t_{s,x})}
{tt_{s,x}| t-t_{s,x}| }
,
\di5
L^Y=\int\limits_{t_1}^{t_2}dt
\frac{u_x(t-t_x)({\cal F}_i(\xi ,t)-{\cal F}_i(\xi ,t_{s}))
+u_s(t-t_s)({\cal F}_i(\xi ,t)-{\cal F}_i(\xi ,t_{x}))}
{(u_s|t-t_s|+u_x|t-t_x|)|t-t_s||t-t_x|}
,
\di5
{\tilde L^k_{s,x}}=
\od1{u_{s,x}}\int\limits_{t_1}^{t_{s,x}} dt
\od {{\cal F}_i(\xi ,t)}{t^2}
-\od1{u_{s,x}}\int\limits_{t_{s,x}}^{t_2} dt
\od {{\cal F}_i(\xi ,t)}{t^2}.
\earr{LLL}
The integration region on variables $\xi=-q^2/2pq$ and $t=-q^2$
($q=k_1-k-k_2$) are plotted on fig.\ref{limint}b. The limits of
integration are
\beq
t_{s,x}=Q^2\od{\{S,X\}}{u_{s,x}},
\qquad
 t_{2,1}={u(S_x\pm\sqrt{\lambda_Q})+2M^2Q^2\over 2(u/\xi+M^2)}.
\eeq{0368}

\subsubsection*{B. Elastic and Quasielastic radiative tails}
In the case of ultrarelativistic approximation in calculation of
RC from elastic radiative tail, due to strong dependence of
formfactors $F_i$ on the square of transfer momenta $Q^2$, the
leading contribution to the total cross section gives only
$t-$peak (Compton peak) and contributions of $s-$ and $p-$ peaks
are suppressed \cite{AKP}. For $\sigma_1^{el}$ we have
\beq
\sigma ^{el}_1=
\sigma ^A_u +
P_LP_N\sigma ^A_p +
\ot {Q_N} 6\sigma ^A_q
,
\eeq{0011}
where index $A$ corresponds to the considered nuclei and $u$, $p$,
$q$ define the unpolarized, polarized and
quadrupolarized contributions. For spin 1/2 nuclei $Q_N=0$ and for
spin 0 nuclei $P_N=Q_N=0$.

For proton, deuteron and carbon we obtain the results
\begin{eqnarray}
\label{0099}
\sigma ^p_u &=&
 \od{\alpha^3}{S} Y_+
\int\limits_{\eta_{min}}^{\infty} \od{d\eta_A}{\eta_A}
  ({\tilde X} (F_1^2 + \eta_A F_2^2) - (F_1+F_2)^2)
,\nonumber \\
\sigma ^p_p  &=&
\od{\alpha^3}{S} Y_-
\int\limits_{\eta_{min}}^{\infty}\od{d\eta_A}{\eta_A} (F_1+F_2)
(x {\tilde X} F_1 - (F_1+F_2))
,\nonumber \\
\sigma ^d_u  &=&
\od{\alpha^3}{S}  Y_+
 \int\limits_{\eta_{min}}^{\infty}\od{d\eta_A}{\eta_A}
  \biggl((F_c^2 + \ot 8 9 F_q^2 \eta_A^2 + \ot 2 3 F_m^2 \eta_A)
{\tilde X}
 - \ot 2 3 (1+\eta_A) F_m^2\biggr)
,\\
\sigma ^d_p  &=&
\od{\alpha^3}{S}  Y_-
\int\limits_{\eta_{min}}^{\infty}\od{d\eta_A}{\eta_A}
F_m\biggl(    \ot 1 2 (1+\eta_A) F_m
- (F_c + \ot 1 3 F_q \eta_A + \ot 1 2 F_m \eta_A) x {\tilde X}
            \biggr)
,\nonumber \\
\sigma ^d_q &=&
\od{\alpha^3}{S}  Y_+
\int\limits_{\eta_{min}}^{\infty}\od{d\eta_A}{\eta_A}
\biggl\{
\Bigl(1+\eta_A+\Bigl(\ot 3 4 x^2 - \eta_A\Bigr) {\tilde X}\Bigr) F_m^2
-\nonumber \\  &&\qquad \qquad  \qquad
-\od {{\tilde X} X_1}{(1+\eta_A)}  F_q  \bigl(3 F_c+3 \eta_A F_m+\eta_A
F_q\bigr)
-\nonumber\\  &&\qquad \qquad \qquad
-2 \eta_A {\tilde X}  F_q  \Bigl(4 F_c-3 x F_m+\ot 4 3 \eta_A F_q\Bigr)
\biggr\}
,\nonumber\\
\sigma^C_u&=&
\od{\alpha^3}{S}Z^2Y_-
\int\limits_{\eta_{min}}^{\infty}\od{d\eta_A}{\eta_A}
{\tilde X}F^2
,\nonumber
\end{eqnarray}
for the case of longitudinally polarized target,
where
\barr{c}
X_1=x^2 + 4 x \eta_A  - 4 \eta_A, \quad {\tilde X}=\od
{X_1}{2\eta_Ax^2}, \quad
Y_{\pm}=\od{1\pm (1-y)^2}{1-y},
\di5
\eta_A=\od{t}{4M^2_A},
\quad \eta_{min}=\od{x^2}{4(1-x)},
\earr{0012}
and formfactors
\beq
F_1={G_E+\eta_A G_M\over 1+\eta_A},\quad
F_2={G_M - G_E\over 1+\eta_A}
.
\eeq{0024}

Polarized contribution in the case of transversely polarized
target is proportional to nuclear mass and therefore equals to
zero in the case of ultrarelativistic approximation \r{apr}.
Nevertheless we present an explicit formulae for the first
nonzero order correction:
\barr{l}
\sigma^p_p   =
\od {\alpha^3 xy^2}{S(1-y)^{3/2}}  \od M Q
\int\limits_{\eta_{min}}^{\infty}\od{d\eta_A}{\eta_A}
\biggl\{
{\cal G}_1^A \biggl({\tilde y_1} \Bigl(2+\od x {y \eta_A}\Bigr)
- \od {(2-y) x} y{\tilde X}\biggr)+
\di5   \;\;\;\;\;\;\;\;\;\;\;\;\;\;\;\;\;\;\;\;\;\;\;\;
+{\cal G}_2^A \biggl(\Bigl(x+2 \eta_A-2 x {\tilde y_1}^2
+6 \od {\eta_A {\tilde y_1}}
y\Bigr)
{\tilde X}  - \od {4{\tilde y_1}}{xy} (1+\eta_A)\biggr)
\biggr\}
,
\earr{poltran}
where ${\tilde y_1}=1/y-1$.
The cross section dependence on nuclei formfactors are contained
in quantities $\cal G$. For protons and deuterons we take
\barr{ll}
{\cal G}_1^p=(F_1+F_2)^2
,\dis
{\cal G}_2^p=F_2(F_1+F_2)
,\di5
{\cal G}_1^d=-\od 1 2 (1+\eta_A)F_m^2
,\dis
{\cal G}_2^d=F_m\bigl(F_c+\ot 13 \eta_AF_q-\ot12F_m\bigr)
.
\earr{0013}
Also we present relations between quadrupolirazation parts for the
cases of longitudinally and transversely polarized targets:
\beq
\sigma^d_{q\bot }=-\ot12
\sigma^d_{q\| }=-\ot12
\sigma^d_{q}
.
\eeq{0114}

For the calculation of the contribution to RC from the
quasielastic radiative tail in the case of ultrarelativistic
approximation we use the results for the proton target
for elastic radiative tail
replacing
\barr{c}
F_e^2(Q^2) \longrightarrow S_e(Q^2)  F_e^2(Q^2),  \quad
F_m^2(Q^2)  \longrightarrow S_m(Q^2)  F_m^2(Q^2),  \di5
F_e(Q^2)  F_m(Q^2)  \longrightarrow S_{em}(Q^2)  F_e(Q^2)  F_m
(Q^2)
.
\earr{supp}

\subsubsection{Higher order effects}                   \label{2.3}
There are no known reasons to consider the
$O(\alpha^2)$ corrections to be negligible. We follow the
method of structure functions used in \cite{KMS} for the
calculation of the higher order electromagnetic radiative
corrections to neutral current unpolarized  lepton-proton DIS
and generalize it for the case of polarized leptons and polarized
nuclei target in the current version of POLRAD.
For the case of $s$- and $p$-peaks the formulae could be
obtained in the terms of the Born cross section and practically
coincide with the expressions for unpolarized particles, however
the contribution of the $t$-peak which is extremely
important in the cases of elastic and quasielastic radiative tails
has to be obtained for polarized DIS.

The sum of second order inelastic correction and
correction due to loop effects has the form
\beq
\sigma ^{in}_2+\sigma^V_2= \delta^{in}_{2}\sigma^0
+\sigma _{Vs}^{in} +\sigma _{Vp}^{in}
+\sigma _{ss}^{in} +\sigma _{pp}^{in}+\sigma _{sp}^{in}
+\sigma _{ls}^{in}+\sigma _{lp}^{in}
+\sigma _{fs}^{in}+\sigma _{fp}^{in}
,
\eeq{0054}
where factorized part is
\barr{c}
\delta^{in}_2=\od
{\alpha^2}{4\pi^2}\left(3\delta^2(Q^2)+
2l_m\delta_{sp}\delta(Q^2)
+\od 1 2 l_m^2
\bigl[ \delta_{sp}^2
+4{\rm Li}_2(1-z_p)+12{\rm Li}_2(1-z_s)-\ot 8 3
{\pi}^2\bigr]\right)
,
\earr{0055}
and $\delta(Q^2)=\delta^l_{vac}+\delta^l_{vac}$.

The contribution from vacuum polarization if coincide with real
photon radiation has the form
\barr{l}
\sigma _{Vs,Vp}^{in}=\od {\alpha^2}{2\pi^2}l_m
\int\limits_{z_{s,p}}^{1}
\frac{dz}{1-z}
\left((1+z^2)\delta(t_{x,s})\sigma _{s,p}-2\delta(Q^2)\sigma_0
\right)],
\earr{l004}
where $t_x=zQ^2$ and $t_s=Q^2/z$.

Next three terms correspond to the cases when two radiated photons
are collinear to incident electron ($\sigma_{ss}$), outgoing
electron ($\sigma_{pp}$)
\barr{l}
\sigma _{pp,ss}^{in}= \od {\alpha^2 }{8\pi^2 }
l_m^2  \int \limits _{z_{p,s}}^1 dz\Biggl [ \od 2{1-z}
(2\ln (1-z)(1-z_{p,s}(z))-\ln z+3)((1+z^2)\sigma _{p,s}-2\sigma
_0)
\di5 \qquad \qquad \qquad
+((1+z)\ln z-2(1-z))\sigma _{p,s}\Biggl]
,
\earr{0056}
or when one photon is radiated in incident electron direction and
the other in the outgoing electron direction ($\sigma_{sp}$):
\barr{l}
\sigma _{sp}^{in}=\od {\alpha^2 }{4\pi^2}l_m^2
 \int \limits _{z_s}^1 \od {dz_1}{1-z_1}
\int\limits _{z_p(z_1)}^1\od {dz_2}{1-z_2}
\Biggl [(1+z_2^2)(1+z_1^2)\sigma _{sp}
-2(1+z_1^2)\sigma _{s}
-2(1+z_2^2)\sigma _{p}
+4\sigma _{0}
\Biggl] .
\earr{00572}
Here
\beq
z_s(z)=(1-y)/(z-xy), \qquad
z_p(z)=(1-y+zxy)/z,
\eeq{00574}
and $\sigma _{s,p}$ depend on $z_{1,2}$ and are given by
\r{0003} with $z\rightarrow z_{1,2}$ respectively. The quantity
$\sigma _{sp}$ depends on both $z_1$ and $z_2$ and also is
calculated in terms of born cross section $\si_0$:
\barr{c}
\sigma _{sp}=y\sigma _0(z_1S,X/z_2,z_1Q^2/z_2)/(z_2(z_2z_1-1+y)).
\earr{997}

There are two channels (singlet and non-singlet) of
fermion pair production that give a contribution to $\alpha^2$
order RC.
The singlet channel
\barr{l}
\sigma _{lp,ls}^{in}=\od {\alpha^2 }{8\pi^2}l_m^2
\int \limits _{z_{p,s}}^{1} dz
(2(1+z)\ln z+1-z+\ot 4 3(1-z^3)z)\sigma _{p,s}
\earr{0058}
corresponds to the case when incident and outgoing lepton as well
as leptons of the unregistered pair belong to
different leptonic lines connected by additional virtual photon.

The rest non-singlet part
\barr{l}
\sigma _{fp,fs}^{in}=\od {\alpha^2 }{12\pi^2 }
\sum_f \ln^2 \od{Q^2}{m_f^2} \int \limits
_{z_{p,s}}^{1-4m_fM/S} dz \od{(1+z^2)}{(1-z)}\sigma _{p,s}
\earr{0060}
arises from the two-lepton decay of additional virtual photon.

The main contribution to second order elastic and quasielastic
radiative tail arises when additionally radiated photon is
collinear to incident or outgoing fermion line:
\barr{c}
\sigma^{el}_2=\od {\alpha }{2\pi}l_m\delta_{sp}\sigma^{el}_1
+\sigma^{el}_{V}+\sigma^{el}_{pt}+\sigma^{el}_{st},
\di5
\sigma^{el}_{s,p\;t}=\od {\alpha }{2\pi}l_m\int
\limits_{z_{s,p}}^1
dz\od {(1+z^2)\sigma^{el}_{s,p}-2\sigma^{el}_{1}}{1-z},
\earr{0118}
where the quantities $\sigma^{el}_{s,p}$ are obtained in terms of
approximate elastic radiative tail \r{0011}
$\sigma^{el}_{1}=\sigma^{el}_{1}(x,y,S)$:
\barr{l}
\sigma _{s}^{el}=y\sigma _1^{el}( xyz/(z+y-1), (z+y-1)/z,
zS)/(z-1+y),
\di5
\sigma _{p}^{el}=y\sigma _1^{el}( xy/(z+y-1), (z+y-1)/z,
S)/z(z-1+y).
\earr{ddfe}
The integral over $z$ in \r{0118} can be calculated
explicitly. For longitudinally polarized target the
results have the  form of eq.\r{0099}
\barr{c}
\sigma ^A_{u,p} =
 \od{\alpha^4l_m}{2\pi S}
\int\limits_{\eta_{min}}^{\infty}\od{d\eta_A}{\eta_A}
\left( {\cal F}_{u,p}^{A1}{\cal R}_1^{u,p} +{\cal
F}_{u,p}^{A2}{\cal R}_2^{u,p}\right)
,
\earr{qq20}
 where the quantities ${\cal R}_{1,2}^{u,p}$ are given in
\ref{appC}. The functions ${\cal F}_{u,p}^{A1,2}$ are quadratic
combinations of nuclear formfactors and could be found by
comparison with \r{0099} which can be written in the
common form
\barr{c}
\sigma ^A_{u,p} =
 \od{\alpha^3}{S}Y_{\pm}
\int\limits_{\eta_{min}}^{\infty}\od{d\eta_A}{\eta_A}
\left( \{{\tilde X},x{\tilde X}\}{\cal F}_{u,p}^{A1} +{\cal
F}_{u,p}^{A2}\right)
.
\earr{qq20p}
The correction due to vacuum polarization $\sigma^{el}_{V}$
is defined by formulae
\r{0011} and \r{0099} with additional factor
${\al \over \pi}\delta(4M_A^2\eta_A)$ under integral.

\subsection{Electroweak radiative correction}          \label{2.4}
The next evident step both from the theoretical and experimental
points of view is the treatment of electroweak effects
contribution. So we included in POLRAD 2.0 the results of
ref.\cite{AISh}
for one-loop electroweak correction within the
framework of standard theory and
the on-shell renormalization scheme in t'Hooft-Feynman gauge.
The result for the correction is obtained as the sum of loop and
radiative effects
\begin{equation}
\sigma_ {1\;ew}^{in} +
\sigma_ {1\;ew}^{v} =
\sigma^B_S+
\sigma_{Vl}+\sigma_{Vq}+\sigma_{box}
+ {\alpha \over \pi} \sum_q \delta_q {\sigma_0^q} +
 \sum_q {\sigma _R^q},
\end{equation}
where $\sigma^B_S$ is the correction to boson propagator,
$\sigma_{Vl,q}$, $\sigma_{box} $ are infrared free parts of lepton
and quark vertex functions and box graphs.
The loop correction is calculated on the basis of
ref.\cite{BHS}.
The quantity $\sigma _R^q$ is an
infrared free part of the real photon emission cross section. The
correction $\delta_q $ is obtained after infrared divergence
cancelation. It is factorized front of the born cross section on
a quark $\sigma_0^q$ and is an analog of quantity
$\delta^{IR}_{R}$ in (\ref{RpV}).
 For radiative effect the methods developed in
ref.\cite{BSh} are applied. The quantity $\sum_q {\sigma _R^q}$
can be derived in terms of leptonic
($\sigma ^{ij}_l$,  ${\hat \sigma }^{ij}_l$), hadronic
($\sigma ^{ij}_h$,  ${\hat \sigma }^{ij}_h$) radiation and their
interference
($\sigma ^{ij}_{lh}$):
\begin{equation}
\sigma _R^q=
\sum_{ij=\gamma ,Z} \left\{
\sigma ^{ij}_l +{\hat \sigma }^{ij}_l
+e_q \sigma ^{ij}_{lh}
+e_q^2\left(\sigma ^{ij}_h + {\hat \sigma }^{ij}_h
\right)
\right\}.
\label{sigrad}
\end{equation}
The quantities
$\sigma ^{ij}_{l,h,lh}$ have the form of one-dimensional integrals
over $\xi$
\begin{equation}
\begin{array}{l}
\displaystyle
\sigma^{ij}_{b}  = {\alpha ^3y \over 4}
\int\limits_x^1 {d\xi \over \xi} \biggl\{R^{ij}_V \biggl[
 T^{ij}_{+b}F^{qij}_V(\xi ) - T^{0ij}_{+b}F^{qij}_V(x)
\biggr]
+R^{ij}_A \biggl[
T^{ij}_{-b}F^{qij}_A(\xi )-T^{0ij}_{-b}F^{qij}_A(x)
\biggr]
\biggl\},
\end{array}
\label{sigs}
\end{equation}
where $T^{ij}_{\pm b}$ are the combinations of
kinematical invariants, $R^{ij}_{V,A}$ and $F^{qij}_{V,A}$ are
functions of electroweak coupling constants and parton
distributions \cite{AISh}. The hat-quantities in (\ref{sigrad})
are not large corrections arising from the non-leading terms of
the expansion of polarization vectors (see \ref {2.1}).

\subsubsection{Correction to leptonic current in QCD-improved
model}
An implementation of QCD-improved parton model for the most
important case of leptonic current correction requires an
additional generalization of (\ref{sigrad}, \ref{sigs})
valid for the simple parton model and cannot be generated directly
because an analytical integration over $Q^2_h$ been already done.

As a result correction takes the same form as eqn.\r{0004}:
\begin{equation}
\sigma_ {1\;lept}^{in}+
\sigma_ {1\;lept}^{v} =
\sigma ^{in}_1+\sigma^v_1= \od {\alpha }{\pi }\delta^{in}_1\sigma_0
+\sigma^V_r
+\sigma _s^{in} +\sigma _p^{in}+\sigma _r^{in}.
\label{eweak}
\end{equation}
The quantities $\sigma _{s,p}^{in}$ and $\delta_1^{in}$ are
defined by \r{sppeak} and \r{0020a}. The Born contribution can be
written as
\beq
\sigma_0=\sum_{i,j=\gamma,Z}\sigma^{ij}_0
=\sum_{i,j=\gamma,Z}\od {\pi \alpha ^2 }{2Q^4}X\{ Y_+R^{ij}_+
F^{ij}_+(x,Q^2)+Y_-R^{ij}_-F^{ij}_-(x,Q^2)\}.
\eeq{778}
Here the following notation is introduced in \r{f1g1ew}.

The contribution of electroweak loops can be found in the form
\beq
\sigma^V_r=\sigma_{0}\bigl(R^{ij}_{\pm}\rightarrow
\delta R^{ij}_{\pm}\bigr),
\eeq{778c}
where
\begin{eqnarray}
 \delta R^{\ga \ga }_{\pm}&=&-2\Pi ^{\ga }
R^{\ga \ga }_{\pm}-2\Pi ^{\ga Z}\chi R^{Z\ga }_{\pm}
\nonumber\di2 &&
+\od{\alpha }{4\pi}[\lambda^{l\ga \ga }_VR^{ZZ}_{\pm}
\Lambda _2(-Q^2,M_Z)
+v^{\ga }_l(1-P_L)\od 3 {s_w^2}\Lambda _3(-Q^2,M_W)],
\nonumber\di5
 \delta R^{\ga Z}_{\pm}=
\delta R^{Z \ga }_{\pm} &=&-2(\Pi ^{\ga }+\Pi ^Z)
(R^{\ga Z}_{\pm}+R^{Z\ga }_{\pm})
-\Pi ^{\ga Z}(R^{\ga \ga}_{\pm}+\chi R^{ZZ}_{\pm})
\nonumber\di2   &&
 +\od{\alpha }{4\pi}[(\lambda^{lZ\ga }_VR^{ZZ}_{\pm}
+\lambda^{l\ga Z}_AR^{ZZ}_{\mp})\Lambda _2(-Q^2,M_Z)
\di2              &&
+(1-P_L)\{\od {v^{\ga }_l}{4s_w^2}\Lambda _2(-Q^2,M_W)
+\od 3{2s_w^2}(v^z_l+a^z_l-\od {c_w} {s_w}v^{\ga }_l)\Lambda
_3(-Q^2,M_W)\}],
\nonumber\di5
\delta R^{ZZ}_{\pm}&=&-2\Pi ^ZR^{ZZ}_{\pm}-2\Pi ^{\ga Z}R^{\ga
Z}_{\pm}
\nonumber\di2      &&
+\od{\alpha }{4\pi}[(\lambda^{lZZ}_VR^{ZZ}_{\pm}
+\lambda^{lZZ}_AR^{ZZ}_{\mp})\Lambda _2(-Q^2,M_Z)
\nonumber\di2        &&
+(1-P_L)(v^z_l+a^z_l)\{\od 1{2s_w^2}\Lambda _2(-Q^2,M_W)
-3\od {c_w}{s_w^3}\Lambda_3(-Q^2,M_W)\}].
\label{778d}
\end{eqnarray}
Here $M_{Z,W}$ are masses of $Z$ and $W$ bosons and
\begin{equation}
\begin{array}{l}
\displaystyle
\Pi ^{\gamma}=-{\hat{\Sigma }^{\gamma }(-Q^2)\over Q^2},\quad
\Pi ^Z=-{\hat{\Sigma }^Z(-Q^2)\over Q^2+M_Z^2}          ,\quad
\Pi ^{\gamma Z}=-{\hat{\Sigma }^{\gamma Z}(-Q^2)\over Q^2}.
\end{array}
\label{Pi}
\end{equation}
Quantities $\hat{\Sigma}^{\gamma, \gamma Z, Z}$ are defined by
formulae (A.2,3.17,B.2-5) of
\cite{Holl} and $\Lambda _{2,3}$ by (B.4,B.6)  of \cite{BHS}.

In the electroweak case the term $\sigma _r^{in}$ has form
\beq
\sigma _{r}^{in} =\sum _{i,j=\ga,Z}
\alpha^3y\int\limits_x^1 \frac{d\xi}{\xi^2}
\left[R_+^{ij}T_+^{ij}+R_-^{ij}T_-^{ij}
+P_L(\la ^{lij}_A\hat{T}_+^{ij}+\la ^{lij}_V\hat{T}_-^{ij})
\right],
\eeq{0009c}
where
\barr{l}
T^{ij}_+=\od {1}{4\xi Q^4_{\xi }} [Q^4_{\xi }
{\cal L}_1( F_+^{ij})+2{\cal L}_2(F_+^{ij})],
\di5
T^{ij}_-=\od{1}{4\xi Q^2_{\xi }}{\cal L}_3(F_-^{ij})
\di5
\hat{T}^{ij}_{\pm}=-{u(S^2\pm u_x^2)\over 2\xi u_xXQ^4_{\xi}}
F^{ij}_{\pm }.
\earr{TTP}
Functions ${\cal L}_{1,2,3}$ are defined in \r{call}.

\subsubsection{Correction to hadronic current}
Exact formulae for correction to hadronic current could be
obtained from \r{sigs} when $b=h$ and could be easily generated
for the case of QCD-improved model (see also ref.\cite{Tim}).

Result for leading log approximation could be presented in
standard form
\cite{ll}:
\begin{equation}
\sum_q e_q^2{\sigma _h^{ij}}=
{\sigma _0^{ij}}\left(f^{\pm }_q(x) \rightarrow
f_q^{\pm rad}(x) \right),
\end{equation}
where
\begin{equation}
\begin{array}{l}
\displaystyle
f_q^{\pm rad} (x)=
e_q^2 \frac{\alpha }{2\pi }  \ln \frac{Q^2}{m_q^2}
\biggl\{
-f_q^{\pm}(x)
\ln{\od{Q^2}{(1-x)^2m_q^2}}
\di5 \qquad\qquad
+\int\limits ^{1}_{x}
\od{dz}{z} \biggl\{
{1+z^2\over 1-z} f_q^{\pm }(x/z)
-{2\over 1-z} f_q^{\pm }(x)
\biggr\}
\biggr\}.
\end{array}
\label{fff}
\end{equation}
Sometimes fits for partonic distributions are constructed from the
data extracted without taking into account
the of hadronic current correction. Therefore if such a fit is
used in the calculation the correction does not have to be taken
into account.

\subsection{Semi-inclusive physics} \label{2.5}

We calculate
the  radiative
corrections to data of semi-inclusive polarized experiments when
a hadron is detected in coincidence with the outgoing lepton. In
this case the cross section depends additionally on variable $z$
defined as
\begin{equation}
z={p_1 p_2 \over p_1 q},
\end{equation}
where $p_1$, $p_2$ ($p_2^2=m_h^2$) and $q$ are 4-momenta of
initial nucleus,
coincident hadron and virtual photon. This variable corresponds
to the amount of virtual photon energy transmitted to measured
hadron in lab frame.

For the Born cross section of semi-inclusive DIS we use the
formula
\begin{equation}
\sigma _{0}^s \equiv {d^{3}\sigma _{0} \over dxdydz} =
{2\pi \alpha ^{2} \over Sxy}[F^{u}_{0} \Sigma ^{+}(x,z) +
P_{L}P_{N}F^{p}_{0} \Sigma ^{-}(x,z)]
\label{71}
\end{equation}
where
\beq
F^{u}_{0} = 2( 1/y - 1 - \mu _{N}x ) + y, \qquad F^{p}_{0} = y -
2, \qquad \mu_N=M^2/S
\eeq{si01}
and $\Sigma ^{+(-)}(x,z)$ are defined in \ref{appA2}.

The lowest order QED correction was
calculated in
ref.\cite{SSh,SSh1} and can be written as the sum of factorizing
and non-factorizing parts
\begin{equation}
\sigma _{EM} \equiv {d^{3}\sigma _{EM}\over dxdydz} =
\sigma ^{F}_{R} +
{\alpha \over \pi } \delta _{VR} \sigma _{0}^s + \sigma_r^V,
\end{equation}
where
\barr{l}
\delta _{VR} = \delta _{V} + \delta ^{IR}_{R} = \delta^s _{inf } -
{1\over 2}\ln ^{2} {r_{3}\over r_{4}} + {3\over 2}l_{m}
- {2} + {\rm Li}_2 (r_{2} / r_{1}) - {\pi ^{2}\over 6},
\di5
\delta^s _{inf } = (l_{m} - 1)\ln (t^{2}_{1i} / r_{1}),
\qquad
t_{1i}=min\{t_{1m},t_{2m}/r_+\}
\earr{si02}
and
\barr{c}
t_{1m}=y(1-x)+\mu_N-(M+m_h)^2/S, \quad
r_{\pm}=[y+2\mu_N\pm
(y^2+4\mu_Nxy)^{1/2}]/2r_9,
\di5
t_{2m}=y(1-z), \quad
r_{1} = 1 - x^{2}y^{2} - yr_{4}, \quad r_{2} = r_{5} - yr_{8},
\quad r_{3}=1-y(1-x),
\di5
r_{4}=1-xy, \quad r_{5}=1-y,
\quad r_8=\mu_Nx,
\quad r_9=\mu_N+y-xy.
\earr{0777}

The finite part of \r{71} has the form
\begin{equation}
\begin{array}{l}
\displaystyle
\sigma ^{F}_{R}
= {2\alpha ^{3}\over S}
\int\limits_{t_{1d}}^{t_{1u}}dt_1
\int\limits_{t_{2d}}^{t_{2u}}dt_2
 \{
{y^{2}\over x^{2}_{t}y_{t}} [F^{u}_{R}\Sigma^{+}
(\tilde{x},\tilde{z}) + P_{L}P_{N}F^{p}_{R}\Sigma ^{-}
(\tilde{x},\tilde{z})] - \\ [0.5cm]
\displaystyle \qquad  \qquad
- {\theta (t_{1i}-t_{1})\over xy} [F^{u}_{IR}\Sigma ^{+}(x,z) +
P_{L}P_{N}F^{p}_{IR}\Sigma ^{-}(x,z)] \},
\end{array}
\label{sigFR}
\end{equation}
where the integration region is defined as (see fig.\ref{limint}c)
\barr{cc}
{t_{1d}}=0, \dis {t_{1u}}=min(t_{1m}, t_{2m}/r_{-}), \di5
{t_{2d}}=t_1 r_-, \dis {t_{2u}}=min(t_{2m}, t_{1} r_{+})
\earr{lim}
and we select the unpolarized $F^{u}_{R} $ and polarized
$F^{p}_{R}$ terms:
\barr{c}
F^{u}_{R} = (x_{t}/y_{t}) \{ [-2\mu I^{2} - I^{1} +
xy\tilde{I}] (r_{5}(1-t_{2}) - \mu _{N}x_{t} + y^{2}_{t}/2) +
\di5
+ [-2\mu \hat{I}^{2} + \hat{I}^{1} + xy\tilde{I}]
(r_{5} + t_{2} -\mu _{N}x_{t} + y^{2}_{t}/2) -
\di5
- (1+yr_{8})I^{1} + (r^{2}_{5} + yr_{8})\hat{I}^{1} - 2\mu _{N}I^{0}
\} + {1\over 2} xy (I^{1} - \hat{I}^{1}) + I^{0},
\earr{si03}
and
\barr{c}
F^{p}_{R} = (1/y_{t}) \{ G[xyy_{t} - 2x_{t}] - \mu y_{t}(t_{1} -
t_{2})\hat{I}^{2} + x^{2}y^{2}t_{2}\tilde{I} +
\di5
+ [xy(y + y_{t}- 1) + (t_{1}-t_{2})(y_{t}/2 -1)] \hat{I}^{1} -
\di5
- [xy(y^{+}_{t}- 1) + (t_{1}-t_{2})(y^{+}_{t}/2-1)] I^{1} \}.
\earr{si04}
In displayed above formulae
\barr{c}
\mu=m^2/S, \quad
y^{+}_{t}=y + t_{2},  \quad
x_{t}=xy+t_{1}-t_{2}, \quad y_{t}=y-t_{2},
\di5
F_{IR}^{\{u,p\}} = F_{0}^{\{u,p\}}G, \quad
G =-\mu (I^{2} + \hat{I}^{2}) + xy\tilde{I}.
\earr{si05}
Here we use the formulae
\barr{c}
I^{0} = \hat{I}^{0} = \Delta ^{1/2},\quad I^{1} = \{
A^{2} + C \}^{-1/2},\quad \hat{I}^{1} = \{ B^{2} + C \}^{-1/2},
\di5
I^{2} = y [(1 + 2r_{8})t_{1} - (r_{7} + xr_{6})t_{2}] (I^{1})^{3},
\di5
\hat{I}^{2} = y [(r_{5}-2r_{8})t_{1} - (r_{5}r_{7} - xr_{6})t_{2}]
(\hat{I}^{1})^{3},
\earr{si06}
where
\barr{c}
A =t_{1} - r_{4}t_{2}, \quad B = r_{5}t_{1} - r_{3}t_{2},
\quad
C = 4\mu\{r_{6}t_{1}t_{2} - \mu _{N}t_{1}^{2} - r_{9}t_{2}^{2}\},
\earr{si07}
and
\begin{equation}
r_{6}=y+2\mu _{N}, \qquad  r_{7}=1-2x.
\end{equation}

To take into account the higher order contribution of soft photons
the standard exponentia\-ti\-on procedure \cite{Sh} is used
\begin{equation}
\delta _{EM} \rightarrow \delta _{exp} \equiv \exp [(\alpha
/\pi)\delta^s _{\inf }] \delta _{EM},
\end{equation}
so that
\begin{equation}
\sigma ^{H} \equiv  \sigma _{0}^s ( 1 + \delta _{\exp }).
\label{sigmaH}
\end{equation}

Since the experimental analysis is often performed
for all pions with $z > z_{0}$ we have to integrate the
expression (\ref{sigmaH}) over $z$. As $z-$dependence is
contained only in quantities $D^{H}_{q}$ and integration limits
\r{lim}, then one can use the following
identity
\begin{equation}
\int ^{1}_{z_{0}} dz \int _{\Omega } dt_{1}dt_{2} \sum _{q}
\varphi_{q} D^{H}_{q}(\tilde{z}) = \int _{\Omega _{0}} dt_{1}dt_{2}
\sum _{q} \varphi _{q} \tilde{D}^{H}_{q}(z_{0},y_{t}/y),
\end{equation}
where
\begin{equation}
\tilde{D}^{H}_{q}(z_{0},\zeta ) = \zeta \int^{1}_{z_{0}/\zeta } dz
D^{H}_{q}(z),
\end{equation}
and $\varphi _{q} \equiv \varphi _{q}(S,x,y,t_{1},t_{2})$ is an
arbitrary function. Hence, substituting some new effective fragmentation
function $\tilde{D}^{H}_{q}(z_{0},y_{t}/y)$ instead of
$D^{H}_{q}(\tilde{z})$ and $\tilde{D}^{H}_{q}(z_{0},1)$ instead of
$D^{H}_{q}(z)$ one can represent the integration over $z$ in
$\sigma ^{H}$  in the form of eq.\ (\ref{sigmaH}) and get
\begin{equation}
\sigma ^{H}(x,y) \equiv \int^{1}_{z_{0}}dz \sigma ^{H}
= \sigma _{0}^s (1 + \delta _{exp}) \left\{
\begin{array}{c}
D^{H}_{q}(\tilde{z}) \longrightarrow \tilde{D}^{H}_{q}(z_{0},y_{t}/y) \\
\Omega \longrightarrow \Omega _{0} \\
D^{H}_{q}(z) \longrightarrow \tilde{D}^{H}_{q}(z_{0},1)
\end{array} \right\} .
\end{equation}

\subsubsection{Experimental cuts}
In the previous section the  integrated
in the whole kinematical region
over hadron variables
$\phi_H$ and $p_{\perp}$ cross
section was considered. For the real situation the
region of integration is limited by the experimental cuts.
To deal with the ones on the angles of registered
hadron a  special procedure was developed.
For cross section we have
\begin{equation}
{d \sigma _{R}
\over dx dy dz}
 = \int
{1\over 2\pi}
 {d^3 k\over 2 k_0}
{d{\tilde z}\over dz}
dp_{\perp}
d\phi_H
\theta({\sin^2 \vartheta_{max}}- {\sin^2 \vartheta})
\theta({\sin^2 \vartheta}-
{\sin^2 \vartheta_{min}})
{d\sigma \over dx dy dz dp_{\perp} d\phi_H},
\label{sigRd}
\end{equation}
where  $\vartheta$ is the angle between the beam direction (${\vec
k_1}$) and hadron momentum in
the lab frame.
One can obtain \r{sigFR} straightway if $\theta-$functions in
\r{sigRd} are removed. We note, that \r{sigRd} is valid when
asimutal symmetry of detector is supposed.
As the distribution on $p_{\perp}$ is unknown it was
approximated by
$\delta$-function
\begin{equation}
D({\tilde z}, p_{\perp})=D({\tilde z})\delta(p_{\perp}){1\over
2\pi},
\label{d}
\end{equation}
according to the normalization condition
\begin{equation}
\int D({\tilde z}, p_{\perp})dp_{\perp}d\phi_H=D({\tilde z}).
\end{equation}
The presence of $\delta-$function in \r{d} takes off the
integration over $p_{\perp}$, so the integration over $\phi_H$
becomes trivial as arguments of $\theta-$functions are now not
dependent on $p_2$ (and on $\phi_H$), but acquires dependence on
photon momenta.

Implying peaking approximation in \r{sigRd} we preserve only $t_1$
dependence for $\theta-$functions, that allow to carry out
analytical integration over one of photon variables. As a
result we have the same formula as \r{sigRd}, but we replace
\barr{c}
I^{1,2}\longrightarrow I^{1,2} \theta (t_1 - r_4(y-d)), \qquad
{\hat I}^{1,2}\longrightarrow {\hat I}^{1,2} \theta \biggl(t_1 -
{r_3(y-d)\over 1-(y-d)}\biggr),
\di5
I^0 \longrightarrow {1\over 2} I^{0} \left\{
\theta (t_1 - r_4(y-d))+
 \theta \biggl(t_1 -
{r_3(y-d)\over 1-(y-d)}\biggr)\right\},
\earr{repl}
where we take into account, that
in nonradiative case the approximation \r{d} is equivalent
to
the assumption, that only those hadrons are registered, for which
\begin{equation}
\beta_{min} \leq {\Delta_3 \over y^2} \leq
\beta_{max},
\label{0020}
\end{equation}
where
\barr{c}
\beta_{min,max}={{\sin^2 \vartheta_{min,max}}\over 2\mu_N},
\qquad d=\left\{ \od{\Delta_3}{\beta_{max}}\right\}^{1/2},
\di5
\Delta_3=y(-4x\mu_N\mu-xy-\mu y  +x-\mu_Nx^2y)
.
\earr{0470}

\subsection{Iteration procedure of data processing}    \label{2.6}

From the POLRAD beginning particular emphasis has been placed on
the procedure of RC of experimental data.
In the current version the iteration procedure,
which
allows to extract Born data sets for cross sections, SF or
asymmetries from observables ones taking into account the
radiative effects is realized both for the cases of inclusive
 experiments.

As an example we consider the procedure of radiative
correction to extract the structure function $g_1(x)$
from measured spin asymmetry
$A_{1i}^m$ with error $\epsilon_i$. The spin average structure
functions are considered to be constant and
$g_{2}(x)$  equals to  0.
The measured asymmetry is defined as
\begin {equation}
A^m_1 = {g_1\over F_1} + \Delta A_1(g_1),
\label{53}
\end {equation}
where
 the radiative correction to asymmetry
$\Delta A_1$
can be written in terms of spin-average and spin-dependent parts
($\si^{u,p}$) of cross sections \r{eq1}
\beq
\De A_1=
{\si_0^u\bigl(\si_p^{in}(g_1)+\si_p^{q}+\si_p^{el}\bigr)
-\si_0^p(g_1)\bigl(\si_u^{in}+\si_u^{q}+\si_u^{el}\bigr)
\over
\si_0^u\bigl((1+\de_v)\si_0^u+\si_u^{in}+\si_u^{q}+\si_u^{el}\bigr)
}.
\eeq{rc02}
 where $\de_v=\si^v_p/\si_0^p=\si^v_u/\si_0^u$. The Born and
inelastic radiative tail polarized parts of cross sections depend
on SF $g_1$, and in the last case the dependence is
non-trivial. So the equation \r{53} becomes functional one in
$g_1$. This functional equation transforms into a system considering the
extraction of $g_1$ in concrete binning over $x$
in $n$ kinematical points $x_i\; (i=1,...,n)$:
\begin {equation}
A^m_{1i} = {g_{1i}\over F_1} + \Delta A_1(g_{1j};j=1,...,n).
\label{54}
\end {equation}
Usually the iteration methods are used to solve such a system of
equations. The variant of iteration formula is ambiguous, but in
practice only two types are used. The first and most evident one
is to take for $k$-th step:
\begin {equation}
g^{(k)}_{1i} = F_1(A^m_{1i} - \Delta
A_1(g^{(k-1)}_{1j};j=1,...,n)).
\label{55}
\end {equation}

Another possibility to obtain the formulae for iteration procedure
arises when both born and radiative correction
cross section
are separated into
spin-averaged and
spin-dependent parts.
Then for the measured asymmetry we have
\begin {equation}
\displaystyle
A^m_1 =\od1D {\sigma_0^p+\sigma_1^p\over
\sigma_0^u+\sigma_1^u} =
{g_1/F_1 + \sigma_1^p/D\sigma_0^u\over
1 + \sigma_1^u/\sigma_0^u}.
\end{equation}
Thus we obtain the iteration formulae
\begin{equation}
g^{(k)}_{1i} = F_1\biggl[A^m_{1i}\biggl(1+\od
{\sigma_1^u}{\sigma_0^u}\biggr)
 - {
\sigma_1^p(g^{(k-1)}_{1j};j=1,...,n)\over D \sigma_0^u} \biggr],
\label{56}
\end{equation}
where in right-hand side the dependence on $g_1$ is contained only
on the level of RC, but not on the Born level.

On the each step of iteration procedure $g^{(k)}_{1i}$ is fitted
with the help of CERNLIB package MINUIT taking into account an
experimental uncertainty $ \epsilon_i (g_1)=
\epsilon_i(A_1^m) F_1$. On the first step we adopt
$g_{1i}^{(0)}=A_{1i}^m F_1$.
The  procedure  converges  within  4-5
steps. As a result we extract values of $g_{1i}$ and
parameters of its fit.

Besides, package MINUIT
is used
to fit the data with the account of
experimental uncertainties  that
gives the opportunity to theoretically calculate the error
propagation of statistical uncertainty of fitted experimental data
to the value of $\Delta A$:
\barr{c}
\ep_i^{ext}=\left[\ep_i^2\bigl(
1+{\bar \delta_{iu}} -{\bar \delta_{ip}} \bigr)^2
+\sum_{j\not= i} {\bar \delta_{ip}}^2\ep_i^2\right]^{1/2},
\di5
{\bar \delta_{iu}}=
\frac{\si^u_{in}+\si^u_{q}+\si^u_{el}}{\si_u^0+\si_u^1}
\sum_{k=1}^{N_p}
\frac{\partial \si_{0}^p}{\partial p_k}
\frac{\partial p_k}{\partial A_j^m},
\di5
{\bar \delta_{iu}}=
\frac{1}{\si_u^0+\si_u^1}
\sum_{k=1}^{N_p}
\frac{\partial \si_{in}^p}{\partial p_k}
\frac{\partial p_k}{\partial A_j^m}.
\earr{er06}
The sum runs over parameters $p_k$ of fitting function $f(x)$.
The first derivative is calculated
 by direct calculation of $\si_p^{in}$ by POLRAD with
using $\frac{\partial f(p,x_i)}{\partial p_k}$ instead of model
$A_1$ (or $g_1$). The second derivative is obtained from the
system of linear equations
\beq
\sum_{k=1}^{N_p}
\frac{\partial p_k}{\partial A_j^m}
\sum_{i=1}^N\frac{1}{\ep_i^2}
\bigl(  A^{m}_{1\; i}
f^{\prime\prime}_{kn}(x_i)
-ff^{\prime\prime}_{kn}(x_i)
-f^{\prime}_{n}(x_i) f^{\prime}_{k}(x_i)
  \bigr)
=-\frac{1}{\ep_j^2}
f^{\prime}_n(x_j)
.
\eeq{er08}

Let us present an explicit formulae for three most important
cases, when proton structure function $g_1$ is extracted from the
data with hydrogen target and neutron structure function $g_1$ is
extracted from the data with deuteron and $^3$He targets. Also in
the last subsection we consider the other target possibilities
without taking into account experimental uncertainties (not using
CERNLIB package MINUIT).

\subsubsection{Proton, Deuteron and Helium-3 targets}
The formulae of previous section could be applied directly for
experimental data on hydrogen and deuteron. Three-parameter fit is
used for fitting of the spin asymmetry \cite{SMC}:
\barr{c}
A_1^p(x)=A+x^B(1-e^{-Cx}),\di3
A_1^d(x)=(e^{-Ax}-1)(B^C-x^C).
\earr{0014}
For proton asymmetry the one- and two parameter fits of
ref.\cite{KNSS} are also used:
\beq
A_1^p(x)=x^B, \qquad
A_1^p(x)=Ax^B.
\eeq{4439}

For  the neutron asymmetry extracted from the deuteron
and helium-3 data the
nuclear corrections have to be taken into account:
\beq A_1^n(x)={A^{D,^3\rm He}_{1}(x)-(1-f_d(x))P_pA_1^p(x)
\over f_d(x)P_n }.\eeq{00d7}

Dilution factors for both cases are given by the formulae
\beq
f_d(x)={1\over F_2^p/F_2^n+1}, \qquad
f_d(x)={1\over 2F_2^p/F_2^n+1},
\eeq{00060}
respectively. The numbers $P_p$ and $P_n$
are effective nucleon polarizations in nucleus. For deuteron
target
$P_p=P_n=0.5-0.75\omega_d$, where $\omega_d$ is the D-state
probability ($\sim$5\%), and
$P_p$=-0.028 and $P_n$=0.86 \cite{Ciofi} for helium-3.

In both cases we construct the fit for neutron asymmetry using
Schaefer's model \cite{Sch} (see Appendix \ref{appA4}).

\subsubsection{Other schemes}
In many experimental cases when the model for structure function
extracted
is unknown it is convenient to use simple spline approximation of
experimental data.
Below we consider two examples, first when $g_1$ and $g_2$ are
extracted simultaneously from data on spin asymmetries $A_1$,
$A_2$ and the second, when $b_1$ is extracted from the data on
quadruple asymmetry $A_q$.

\begin{equation}
\begin{array}{l}
   \displaystyle
A_1^{m}=A_1^{B}(x)+\Delta A_1(g_1,g_2)={g_1-\gamma ^2 g_2\over
F_1}+\Delta A_1(g_1,g_2),
  \\[0.5cm] \displaystyle
A_2^{m}=A_2^{B}(x)+\Delta A_2(g_1,g_2)={\gamma (g_1+ g_2)\over
F_1}+\Delta A_2(g_1,g_2),
\end{array}
\end{equation}
where $\gamma={\sqrt Q^2}/\nu$.
SF $g_1(x)$ and $g_2(x)$ are calculated on the each step:
\begin{equation}
\begin{array}{l}
     \displaystyle
g_1^{(n)}={F_1 \over (1+\gamma ^2)}\left( A_1^{m}-\Delta
A_1(g_1^{(n-1)},g_2^{(n-1)})
+\gamma (A_2^{m} - \Delta A_2(g_1^{(n-1)},g_2^{(n-1)}))\right),
     \\[0.5cm] \displaystyle
g_2^{(n)}={F_1 \over \gamma (1+\gamma ^2)}\left( A_2^{m}-\Delta
A_2(g_1^{(n-1)},g_2^{(n-1)})-\gamma (A_1^{m}
- \Delta A_1(g_1^{(n-1)},g_2^{(n-1)}))\right) .
\end{array}
\end{equation}

For quadruple deuteron spin-dependent SF $b_1(x)$
we have:
\begin{equation}
A_q^{m}={b_1\over F_1} +\Delta A_q(b_1)
\end{equation}
and
\begin{equation}
b_1^{(n)}=F_1 \left( A_q^{m}-\Delta A_q(b_1^{(n-1)})\right) .
\end{equation}

Also the spline method, described above can be used for the cases
of previous section to obtain the model independent data.

\section{User manual}

\subsection{Program structure}

In this section we present the common structure of POLRAD 2.0
along with the short description of subroutines used.

\subsubsection{Main program body}

The code POLRAD 2.0 operates under PATCHY from CERNLIB
\cite{YPAT}. It means
that there are two files POLRAD20.CAR and POLRAD20.CRA. The second
file contains a set of switches for user. The first one is text of
code for all possible combinations of the switches. FORTRAN code
is obtained by calling like {\it pat polrad20}, where {\it pat} is
a simple executable file. For examle for SUN station it can have
form:
\begin{verbatim}
ytobin $1 $1 $1 $1 .go
ypatchy $1 $1 $1 $1 .go
rm y.lis
rm $1.pam
rm $1.lis
\end{verbatim}

POLRAD 2.0 allows user to conduct the calculations and to obtain
the results for the any set of the
the following positions (in brackets we refer to the corresponding
theoretical description).

1) Exact calculation of the total RC to DIS of polarized particles
(see subsection \ref{2.1});

2) Electroweak RC calculation along with 1) (see subsection
\ref{2.4});

3) The contribution of $\alpha^2$ order correction
calculation along with 1) (see subsection \ref{2.3});

4) Approximate calculation of the total RC
(see subsection \ref{2.1});

5) Exact calculation of the total RC to SIDIS of polarized
particles.

Hence, the main file POLRAD20.CAR
consists from the parts
(patches) POLRAD; EXACT; POLRAD\_ADD; SIRAD corresponding an above
mentioned positions, parts (patches) STRFFUN; FITS2; INTEGRAT
of the common use.
The correlation between the patches is presented on the
figs.\ref{schemPOL}-\ref{schemGWS}.

\noindent{\bf Patch POLRAD}.
POLRAD guides the calculation and sends the obtained data in output
files.
On the first step of program run subroutine\footnote{Usually one
subroutine corresponds to one DECK from PATCHY with the same
name.} CONKIN sets up the value of
the invariants in dependence on the given lepton, target and kinematics.
Subroutines FSPENS and FSPEN calculate Spence function.
Subroutine DELTAS calculates the factorizing part of virtual and real
leptonic bremsstrahlung. Subroutine BORNIN calculates the Born
cross section on the basis of known formulae \r{bt}.

\noindent {\bf Patch EXACT}.
In TAILS and FFU the $\tau$-dependent part of an integrand
($\theta _{ij}(\tau)$ in (1)) is calculated.
The integration procedure has the following steps:

1) QQT and QQTINT set up the integration parameters and call the
integrators;

2) Integrators use the  integrands which
are calculated in subroutines RV2DI, RV2 and PODINL.

\noindent{\bf Patch SIRAD} deals with semi-inclusive RC. Main
program MAINPRG guides the calculation of Asymmetry or $r(z)$.
     Subroutine COMVAR defines some often used variables and
     the limits for the phase space of the emitted photon.
     FSP and FSP1 give the Spence function.
 Subroutines
 EXHH, QXT, INTEG and NII carry out integration over photon
variables.
Models for parton distributions and fragmentation functions are
given in DZ, FITDZ, SIGMA, QS QP.
     Subroutine DOS and DOP defines unpolarized and polarized
Born cross section.

     Functions FYS(F0YS) and FYP(F0YP)  define the sum and
difference of cross sections with the target
     polarization vector parallel and antiparallel to the momentum
     vector of the incident lepton. Radiative effects are
(are not) included.

     Functions FCS(FCP) and FRS(FRP) are the integrands of the
unpolarized (polarized)
     part of bremsstrahlung cross section over invariant
     variable corresponding to the energy and to the polar angle
of the emitted
     photon.

 Subroutines VPQRK and Z0\_EXH calculates
 effects contribution of the quarks to the vacuum polarization and
effects of $Z_0$-exchange at the Born level.

 Subroutines INPUT and OUTPUT create input arrays of
kinematics variables and form output files.

\noindent{\bf Patch STRFFUN}.
Subroutine STRF carries out the calculation of inelastic, elastic
quasielastic SF ($\Im_{i}(R,\tau)$), defined for different cases
in \ref{appA} as the combinations of usual unpolarized
($F_2$, $F_1$ or $R$), polarized ($g_1$, $g_2$), quadruple
($b_{1-4}$) structure functions and formfactors.

SF $F_2$ is calculated in dependence on the user defined model for
the whole kinematical range in the subroutines:

COMFST and F2BRAS
calculate $F_2$ outside and inside resonance region on the basis
of experimental data from refs. \cite{Ste,Bra,Mil};

PGRV corresponds to the calculation of SF $F_2$ in according to
the ref.
\cite{GRV} model.

Subroutines DF2H8 and DF2D8 calculate $F_2$ for proton and
deuteron within the ref. \cite{D8} model.

RANUCL gives the relation between $F_2$ for deuteron and $F_2$ for
carbon and oxygen;

F2SFUN is subroutine managing the calculation of $F_2$.

Subroutine R1990 calculates R from ref.\cite {Whi}.

$F_1$ is calculated in subroutine F1SFUN.

For the calculation of $g_1$ in different models,
the following subroutines can be used:

PARPOL calculates $g_1$ in ref. \cite{GRSV96} model;

SCHAF corresponds the calculation of $g_1$ using ref. \cite{Sch}
model;

G1SFUN is subroutine managing the calculation of $g_1$.

$g_2$ equals to zero or is calculated in subroutine G2SFUN within
the model of ref. \cite{WW}.

Quadruple structure functions are calculated in the subroutine
B14SF.

Subroutines FFPRO, FFDEU, FFHE3 and FFCO correspond the
calculation of proton, deuteron, $^3\rm He$, carbon and oxygen
formfactors.

Subroutine FFQUAS calculates formfactors taking into account
quasielastic suppression.

SUPST calculates suppression factors as in ref.\cite {Ste};

PORTN joins together models for SF calculations in
different kinematical regions. As a result the
continuous fit for all SF in whole kinematical region is obtained.

\noindent{\bf Patch FITS2}
contains subroutines for fit modeling based on the different methods:

AMNK, GRAM, GAUSS, FI, BASS --- on the method of the minimal
squared deviations;

ADIDI, DIVDIF --- on the Newton interpolation method;

COEFSP --- on the cubic spline method;

MINSTA --- on the method of minimal squared deviation by MINUIT
(requires CERNLIB).

Subroutine REMNK2 reads a data file, calls fitting subroutines
and stores the parameters of constructed fits.
The constructed fit can be called by subroutine FITFUN in any place of
POLRAD.

\noindent{\bf Patch INTEGRAT.}
This file contains standard integrators D01FCE,
from FORTRAN library NAGLIB which are used for the
double integration over the square region. Single integration is
carried out by QUNC8 by the Newton-Cotes method of 8th order,
DQG32 by the Gauss method or  SIMPS by Simpson method.

\noindent{\bf Patch POLRAD\_ADD} corresponds to the calculation of
electroweak
RC calculation, $\alpha^2$ order calculation and approximate RC
calculation. It consists on the set of the subroutines (decks),
that can be divided in three parts: 1)~subroutines managing the
calculation, 2)~subroutines that are integrands and
3)~auxiliary subroutines.

1)  APPTAI manages approximate and electroweak (only if partonic
distributions are
$Q^2$ dependent) calculations;

TARGWS and GWS manage electroweak corrections calculation when
partonic distributions are $Q^2$ independent;

AL2LL manages the calculation of $\alpha^2$ corrections
contribution;

2)  Here we present the list of subroutines, which calculate the
integrands of the corresponding formulae:
PEAK1        \r{sppeak},
PEAK2 \r{sppeak},
UPRE (\ref{0009a},\ref{0009b}),
ELU \r{0099},
ELP (\ref{0099},\ref{poltran}),
ELQ \r{0099},
RA2IPP \r{0056},
RA2ISS \r{0056},
RA2ISP \r{00572},
ELUAL2 \r{qq20},
ELPAL2 \r{qq20},
RA2LSS \r{0058},
RA2LPP \r{0058},
RA2FSS \r{0060},
RA2FPP \r{0060},
FXI \r{sigs}.

3) SIGMAB and BOURSC  calculates the quantities
(\ref{0003},\ref{997}) and
(\ref{bl},\ref{br}).
 VERCON define electroweak coupling constants;
SIGALL, VERTS and TT5 calculates one loop effects: polarization
operators \r{Pi}, vertex functions and boxes (eqs.(10) and (12) of
ref.\cite{AISh}) respectively.
 FHOLL and DLAMB are auxiliary function defined in (B.1,B.4,B.6)
of ref.\cite{BHS}
VCONEW calculates born cross section with and without taking into
account loop effects.
 FFVAPM calculates functions \r{f1g1ew}.
DISTR calculates partonic distributions.
STRFP2 is auxiliary subroutine corresponding to the calculation of
structure functions.

\subsubsection{Input, Output files}

User can set input parameters for POLRAD 2.0 run in files
POLRAD20.CRA, INPUT.DAT and ITDAT*.DAT.

\noindent{\bf Input file POLRAD20.CRA} contains switches to get
necessary case of a calculation. Below switches are given
with short comments.

This group of switches corresponds the names of patches, which
have to be
set on if the correspondent part of the correction is calculated:
\begin{quotation}

{\it polrad} - switch on inclusive DIS run of POLRAD ;

{\it polrad\_add} - gives the opportunity
to calculate the electroweak RC, $\alpha^2$
order effects and QED RC by approximate methods;

{\it sirad} - switch on SIDIS run of POLRAD;

{\it strffun} - calculates SFs (always on);

{\it fits2} - launches the fitting procedure;

{\it integrat} - adds subroutines for integrations;

{\it exact} - adds subroutines for exact calculation of RC to
DIS.
\end{quotation}

The next switches have to be used to choose the way of
calculations, type of leptons and target polarization:

\begin{quotation}

{\it approx} -  gives opportunity to use approximate
methods together or instead of exact calculation of RC
to DIS;

{\it alpha2ll} - calculates the $\alpha ^2$ order correction in
leading log approximation;

{\it elect} - switches on the electron as an input lepton;

{\it muons} - switches on the muon as an input lepton;

{\it long} -  switches on
the longitudinal polarization of a nuclear
target;

{\it tran} -
switches on
the transverse polarization of a nuclear
target.
\end{quotation}

The next switches are necessary for selection of kinematics within
iteration procedure. In this case the input data are taken from
ITDAT*.DAT.

\begin{quotation}

{\it iter\_pr} - launches the iteration procedure for $g_1$ or
$b_1$;

{\it minuit} - sets the fitting of experimental data by MINUIT
(only in the case of iteration procedure run for $g_1$);

{\it err\_prop} - calculates error propagation factor in
according to eq.\r{er06}.

{\it iter\_pr\_g2} - launches the iteration procedure for $g_2$.
\end{quotation}

The next switches are necessary for selection of kinematics beyond
iteration procedure. The only one has to be set on. It should be
noted that user can also specify kinematics in the beginning of
main program body.

\begin{quotation}

{\it kin\_net} - switches on the kinematical net over $x$, $y$.

{\it kin\_smc} - switches on the SMC kinematical set;

{\it kin\_hermes} - switches on the HERMES kinematical set;

{\it kin\_e142} - switches on the E142 kinematical set;

{\it kin\_own} - switches on the user defined kinematical set (see
file INPUT.DAT description).
\end{quotation}

The following switches allow to select the type or nuclear target.
 The only one has to be set on.

\begin{quotation}

{\it targ\_h} - switches on proton target;

{\it targ\_d} - switches on deuteron target;

{\it targ\_he3} - switches on Helium-3 target;

{\it targ\_c} - switches on carbon target;

{\it targ\_o} - switches on oxygen target.
\end{quotation}

The next keys allow to select models for SF, partonic
distribution and fragmentation function calculation
(see \ref{appA3} also). They correspond to

\begin{quotation}

{\it f2nmc\_d8} -  the model for $F_2$ from  ref.
\cite{D8};

{\it f2comfst} -   the model for $F_2$ from  ref.
\cite{Bra,Ste,Mil};

{\it f2g1sch} -
 the model for $F_2$ and $g_1$ from  ref.
\cite{Sch};

{\it f2g1grsv96} -   the model for $F_2$ and
$g_1$ from  ref. \cite{GRSV96};

{\it f1qpm} -   the Callan-Gross relation for
$F_1=F_2/2x$;

{\it r\_eq\_0} -  $R=0$;

{\it g1asym} -
the
calculation of $g_1$ when iteration procedure is
off
and switch f2nmc\_d8 or f2comfst is set on;

{\it qdstr\_gu} -
 the model of ref.\cite{GU} for partonic
distributions;

{\it g2\_eq\_0} -  $g_2=0$;

{\it g2\_ww} -  the model for $g_2$ from ref.\cite{WW};


{\it ffrg\_aub} - the
  model for fragmentation function from ref.\cite{AUB};

{\it ffrg\_cmb} -  the
 model for fragmentation function from ref.\cite{CMPB};

{\it ffrg\_arn} -  the
model for fragmentation function from ref.\cite{ffarn}.
\end{quotation}

The next switches allow to specify the quantities to be calculated:

\begin{quotation}

{\it born} - the only Born DIS cross section is calculated;

{\it pol\_asym} - polarized and unpolarized parts of the total
cross section are calculated separately;

{\it qua\_asym} - quadruple and unpolarized parts of the total
cross section are calculated separately;

{\it cr\_sec} - total cross section is calculated for all
polarized state;

{\it onlyin} - switch to exclude the contribution of elastic and
quasielastic tails to RC;

{\it output\_a} - sets asymmetry as a SIRAD output;

{\it output\_r} - sets quantity $r(z)$ as a SIRAD output;

{\it intdy} - sets the integration of cross section over $y$;

{\it intdz} - sets the integration of cross section over $z$;


{\it cuts} - applies the kinematical cuts for RC in
semi-inclusive case.
\end{quotation}

The next keys guides the calculation of electroweak effects:

\begin{quotation}

{\it electroweak} - calculates  the total radiative
correction including electroweak RC using partonic distributions;

{\it ew\_onlyqed} - calculates the electromagnetic
RC using partonic distributions by eq.\r{sigs};

{\it ew\_onlylep} - calculates the
RC to leptonic current using partonic distributions by eq.\r{sigs};

{\it eweak} - calculates the BORN electroweak RC to SIDIS;
\end{quotation}

The last group of keys serves for selection of final hadron
type in the case of SIDIS. They switch on the following particles
as registered hadron for SIDIS:

\begin{quotation}

{\it proton} -  proton;

{\it a\_proton} -
antiproton;

{\it k\_minus} -
$K^-$;

{\it k\_zero\_bar} -
${\bar K^0}$;

{\it k\_plus} -
$K^+$;

{\it k\_zero} -
$K^0$;

{\it pi\_minus} -
$\pi^-$;

{\it pi\_plus} -
$\pi^+$;

{\it pi\_zero} -
$\pi^0$;

{\it pi\_diff} -
the difference between $\pi^+$ and
$\pi^-$ production to be the measured observable for SIDIS.
\end{quotation}

\noindent{\bf Input file INPUT.DAT} contains eight lines for
DIS
and nine lines for SIDIS cases. They correspond to lepton
momentum,
target momentum, lepton and nucleous polarization degrees,
quadrupolarization degree (for deuteron target). The rest
lines are used only for {\it kin\_own} switch: number of ($x$,
$y$)
pairs for DIS (triads ($x$, $y$, $z$) for SIDIS) and arrays of
these pairs (or triads).

\noindent{\bf Input file BRASSM.DAT}
gives coefficients of Brasse ref.\cite{Bra} for construction of SF fits
in
resonance region.

\noindent{\bf Input file BB1FIT.DAT}
gives coefficients for construction of quadruple SF $b1(x)$ fit
\cite{KH}.

\noindent{\bf Input file STDLOA1.GRI}
is used for the calculation by model of polarized partonic
distributions of ref.\cite{GRSV96}.

\noindent{\bf Input-output files ITDAT1.DAT, ITDAT2.DAT,
ITDAT3.DAT, ITQUAD.DAT, ITASM2.DAT} contain input-output
information for iteration procedure for the cases of extraction
of $g_1^p$, $g_1^d$, $g_1^{^3 \rm He}$, $b_1$ and $g_2$ respectively.
All of them are organized in the same way. Each line if it is not
a comment (a symbol (not '0') in the first position) gives the
information on the one kinematical point to be processed: $x$,
$y$ (or $-Q^2$ if negative),
measured asymmetry, last step extracted asymmetry,
previous step extracted asymmetry, measured error, output factor
for error recalculation given by \r{er06}.

It should be noted that a kinematical point can be removed from
analysis if to comment this line typing a symbol in the first
position. If the symbol is zero correspondent point is processed
but does not participate in fitting procedure.

\noindent{\bf Output files ALL.DAT, ASM.DAT, TAILS.DAT} have a
title lines with information about version,  switches used
for the calculation in files POLRAD20.CRA and INPUT.DAT.

The file ASM.DAT gives for each calculated kinematical point
quantities $x$, $W^2$, $Q^2$, Born, observed asymmetries and
difference between them.

The file ALL.DAT gives a technical information for each
kinematical
point about quality of numerical integration of all tails and
their
parts separately for polarization and unpolarization parts of
cross section.

The file TAILS.DAT gives eight terms contributed to
radiative correction to asymmetry \r{rc02}.

{\bf Output files ALLP.DAT, ALLU.DAT} are usually used to
plot a
different output quantities.
Each line of these files gives the quantities $x$, $y$, $s$,
$\si_0^{u,p}$,
$\si^{u,p}$,
$\de_1^{in}$,
$\si^{in\;u,p}$,
$\si_2^{in\;u,p}$,
$\si_{1\;ew}^{in\;u,p}$,
$\si^{el\;u,p}$,
$\si_2^{el\;u,p}$,
$\si_{1\;ew}^{el\;u,p}$,
$\si^{q\;u,p}$,
$\si_2^{q\;u,p}$,
$\si_{1\;ew}^{q\;u,p}$ respectively.

It should be noted when the electroweak effects are not
calculated the approximate results (if {\it approx} is set on) for
$\si_{1}^{u,p}$ are printed instead of
$\si_{1\;ew}^{u,p}$.

{\bf Output files DATA.DAT,
FIT.DAT} are intended for the
control of
fitting quality. First file contains base points for the fit and
the second contains points of the fitting curve.

\subsection{Some examples}

In this subsection we give three basic examples to illustrate the
POLRAD 2.0 run: iteration procedure of data processing with
MINUIT in the case of helium-3 target; the calculation of
the radiative correction factor for collider DIS;
 semi-inclusive RC with and
without applying kinematical cuts.

\subsubsection{Iteration procedure with random input}
\label{exampleit}
Here we demonstrate the RC procedure run within experiments on DIS
with polarized Helium-3 target with HERMES beam energy
$E=27.5$GeV. Values of measured asymmetry
as well as averaged $x$ and $Q^2$ were obtained randomly. The
following switches were active:
{\it polrad},
{\it strffun},
{\it integrat},
{\it fits2},
{\it exact},
{\it elect},
{\it long},
{\it iter\_pr},
{\it minuit},
{\it targ\_he3},
{\it f2nmc\_d8},
{\it g2\_eq\_0},
{\it pol\_asym}.

The output file ITDAT3 after one step of iteration procedure is
given in Appendix \ref{appD1}.
Fig.\ref{fex1} shows these results together with constructed fit
of neutron asymmetry (see Appendix \ref{appA4}).

\subsubsection{Radiative correction in experiments at collider}

Here we give results for unpolarized and polarized radiative
correction factors
\begin{equation}
 \delta_u=\frac{\sigma^u_{0}+\sigma^u_{1}+\sigma^u_{2}} {\sigma^{u}_0},
\qquad
 \delta_p=\frac{\sigma^p_{0}+\sigma^p_{1}+\sigma^p_{2}} {\sigma^{p}_0}
\label{RCF}
\end{equation}
within kinematics close to future polarized experiments at HERA
collider (lepton and proton beam energies equal to 27.5 GeV and
810 GeV respectively).
 The following switches were active:
{\it polrad},
{\it strffun},
{\it integrat},
{\it polrad\_add},
{\it fits2},
{\it alpha2ll},
{\it exact},
{\it elect},
{\it long},
{\it kin\_net},
{\it targ\_h},
{\it f2g1grsv96},
{\it g2\_eq\_0},
{\it pol\_asym},
{\it onlyin},
{\it electroweak}.
 The quantities (\ref{RCF}) are shown in fig.\ref{ex2}.
The output file ASM.DAT is
given in Appendix \ref{appD2} (the only $x=.001$ case is kept).

\subsubsection{Semi-inclusive radiative correction with and without cuts}

Here we illustrate the run of the code for semi-inclusive DIS.
 The following switches were active:
{\it sirad},
{\it strffun},
{\it integrat},
{\it elect},
{\it kin\_own},
{\it targ\_h},
{\it qdstr\_gu},
{\it splineff},
{\it ffrg\_aub},
{\it outfun\_a},
{\it intdz},
{\it cuts},
{\it pi\_diff}.

The fig.\ref{ex3} shows Born $A^{born}$ and observed
$A^{obs}$ asymmetry
as well as relative correction
$\de=(A^{obs}-A^{born})/A^{born}$
 with and without taking
into account kinematical cuts.
The output files ASM.DAT with and without applying of
experimental cuts
are given in Appendices \ref{appD3a} and \ref{appD3b}.

\section{Tests and implementation of POLRAD} \label{6.0}
POLRAD passed a number of both analytical and numerical tests. It
was shown that the combinations of coefficients $\Theta(\tau)$
coincide with the corresponding combinations of coefficients from
ref. \cite{MT}.
Also with the help of the algebraic programming system REDUCE 3.5
we proved that the formulae for electroweak inelastic unpolarized
radiative correction are the same as in Appendix C
of ref.\cite{Bardin}.
The spin-independent part of POLRAD was tested numerically in
comparison with
FORTRAN codes TERAD86, FERRAD35 and HECTOR \cite{HECTOR} and
revealed an
excellent agreement practically in all kinematical regions when all
input models and parameters were the same.
The spin-dependent part of POLRAD was compared with the program
of Kukhto and Shumeiko \cite{KSh} and
with
E143 radiative correction program kindly placed in our disposal
by Linda Stuart \cite{Linda}. We found the satisfactory agreement
between all
three programs when choosing the same models and parameters again.
It was impossible to test the part of POLRAD corresponding to the
calculation of spin one particle quadruple polarization due to the
total absence of the results in this region.

Also POLRAD is self-tested program: the part corresponding to
approximate calculati\-on (see Sect. \ref{2.2}) have excellent
agreement
with the exact results; the QED part of electroweak correction
(Sect. \ref{2.4}) coincides with the calculations by the
corresponding exact formulae for the QED lowest order correction
(\ref{RpV}); under the simple modifications of the
semi-inclusive
formulae, the corresponding results coincide with the inclusive
ones.

Now POLRAD is used as the basic and official program for the
procedure of radiative corrections in SMC (CERN) and HERMES
(DESY) and together with the above menti\-o\-ned E143 program in
SLAC experiments with polarized particles.


\appendix
\renewcommand{\thesection}{Appendix \Alph{section}}
\renewcommand{\thesubsection}{\Alph{section}.\arabic{subsection}}
\renewcommand{\theequation}{\Alph{section}.\arabic{equation}}
\section{Structure functions}
\setcounter{equation}{0}

\label{appA}

\subsection{Inclusive structure functions}

\label{appA1}

For hadronic tensor  we use
\begin {equation}
\begin {array}{l}
\displaystyle
W_{\mu \nu }= - \tilde{g}_{\mu \nu }\Im _{1}
+ {\tilde{p}_{\mu }\tilde{p}_{\nu }\over M^{2}}\Im _{2}
+i{
 \epsilon _{\mu \nu \alpha \beta }q_{\alpha }\eta _{\beta }
\over M}\Im _{3}
-i{
 \epsilon _{\mu \nu \alpha \beta }q_{\alpha }p _{\beta }
 (q\eta) \over M^{3}}\Im _{4} \\[0.5cm]
\displaystyle
\;\;\;\;\;\;\;\;\;+ \tilde{g}_{\mu \nu }k_{n}\Im _{5}
- {\tilde{p}_{\mu }\tilde{p}_{\nu }\over M^{2}}k_{n}\Im _{6}
- (\tilde{g}_{\mu \nu }+3\tilde{\eta}_{\mu }\tilde{\eta}_{\nu})\Im _{7}
- {3\over 2} \tilde{\Omega }_{\mu \nu }\Im _{8},
\end {array}
\label{ht}
\end {equation}
where

\begin {equation}
\begin {array}{cc}
\displaystyle
k_{n}={3(q\eta )^{2}-Q^{2}\over M^2},  & 
\displaystyle
\tilde{\Omega } _{\mu \nu }={\pmatrix{\tilde{p}_{\mu }\tilde{\eta }_{\nu
 }+\tilde{\eta }_{\mu }\tilde{p}_{\nu }} q\eta \over M^{2}}, \\[0.5cm]
\displaystyle
\tilde{g}_{\mu \nu }=g_{\mu \nu } + {q_{\mu }q_{\nu }\over Q^{2}} ,
& \displaystyle
\tilde{p}_{\mu }(\tilde{\eta }_{\mu })= p_{\mu }(\eta _{\mu })+ {pq
(\eta q)\over Q^{2}} q_{\mu }.
\end {array}
\label{eq9}
\end {equation}

    The quantities $\Im  _{i}$ are defined  as some combinations  of
SF and formfactors, and the exact expressions in dependence of the
variant of calculation  are considered below in the correspondent
sections.
The dependence of the hadronic  tensor on $pq$ and polarization  degrees
is included in  $\Im _{i}$ too.

We note, that the hadronic tensor for spin-1/2 particle
is derived from (\ref{ht})
by  putting $Q_{N}=0$. The  formulae  for  the  covariant
representation of polarization vector are also valid in  this
case. The hadronic tensor for scalar particles is derived by
$P_{N} = Q_{N} = 0$.

Defining $\epsilon  =M^{2}/pq$ we have
for various $\Im _{i}$ in the case of IRT

\begin {equation}
\begin {array}{ll}
\Im _{1}= F_{1}+ {Q_{N}\over 6} \;b_{1} ,& 
\Im _{2}= \epsilon \pmatrix{F_{2}+{Q_{N}\over 6}\;b_{2}} ,\\[0.5cm]
\Im _{3}= P_{N}\;\epsilon \pmatrix{g_{1}+g_{2}} ,& 
\Im _{4}= P_{N}\;\epsilon ^{2}\;\;g_{2} ,\\[0.5cm]
\Im _{5}= {Q_{N}\over 6}\;\epsilon ^2\;\;b_{1} ,& 
\Im _{6}= {Q_{N}\over 6}\epsilon ^{3}\pmatrix{b_{2}/3+b_{3}+b_{4}}
 ,\\[0.5cm]
\Im _{7}= {Q_{N}\over 6}\epsilon \pmatrix{b_{2}/3-b_{3}} ,& 
\Im _{8}= {Q_{N}\over 6}\epsilon ^{2}\pmatrix{b_{2}/3-b_{4}} .
\end {array}
\end {equation}

Definitions of
SF $F_i, g_i$ and $b_i$ are the same as in ref. \cite {HJM}.

Explicit form  of  the
expression for elastic nuclear formfactors depends on a target
spin.
 For deuteron we have
\begin {equation}
\begin {array}{l}
\Im ^{el}_{1}= {1\over 6} \eta_A F^{2}_{m}(4(1+\eta_A)+\eta_A Q_N),\\[0.5cm]
\Im ^{el}_{2}=\pmatrix{F^{2}_{c}+{2\over 3}\eta_A F^{2}_{m}+{8\over 9}
    \eta_A^2F^{2}_{q}} \\[0.5cm]
\;\;\;\;\;\;\;\;\;\;    +{Q_N\over 6}\pmatrix{\eta_A F_m^2+
    {4\eta_A^{2}\over 1+\eta_A}({\eta_A\over 3}F_q+F_c-F_m)F_q} ,\\[0.5cm]
\Im ^{el}_{3}= -{P_N\over 2}(1+\eta_A)F_m({\eta_A\over 3}F_q+F_c),\\[0.5cm]
\Im ^{el}_{4}= {P_N\over 4}F_m({1\over 2}F_m-F_c-{\eta_A\over 3}F_q ),\\[0.5cm]
\Im ^{el}_{5}= {Q_N\over 24} F^2_m ,\\[0.5cm]
\Im ^{el}_{6}= {Q_N\over 24} \pmatrix{F_m^2
    +{4\over 1+\eta_A}({\eta_A\over 3}F_q+F_c+\eta_A F_m)F_q} ,\\[0.5cm]
\Im ^{el}_{7}= {Q_N\over 6}\eta_A(1+\eta_A)F^{2}_{m} ,\\[0.5cm]
\Im ^{el}_{8}= -{Q_N\over 6}\eta_AF_{m}(F_m+2F_{q}).
\end {array}
\label {chana}
\end {equation}
Here $\eta_A =t/4M^{2}_{A}=(Q^2+R_{el}\tau)/4M^{2}_{A}$,
  and $F_{c}, F_{m}, F_{q}$ - charge,
quadruple formfactors of deuteron.

We also give the expressions for  the  case
of arbitrary spin-1/2 nuclei
\begin {equation}
\begin {array}{ll}
\displaystyle
\Im ^{el}_{1}=  Z^{2}\eta_A G^{2}_{m},& 
\displaystyle
\Im ^{el}_{2}=  Z^{2} {G^{2}_{e} + \eta_A G^{2}_{m}\over 1+\eta_A},\\[0.5cm]
\displaystyle
\Im ^{el}_{3}=  {P_NZ^{2}\over 2} G_{m}G_{e},& 
\displaystyle
\Im ^{el}_{4}=  {P_NZ^{2}\over 4} G_{m}{G_{e} - G_{m}\over 1+\eta_A},
\end {array}
\label {ff}
\end {equation}
and for scalar nucleus
\begin {equation}
\Im ^{el}_{2}=Z^{2} F^{2},
\label {f0}
\end {equation}
where $Z$ is the nucleus charge. All but indicated SF must be  set
equal to zero.

The quantities $\Im ^q_i$ can be obtained in the terms of quasielastic
response functions, which have a form of peak for $\omega =Q^2/2M$. The
fact is normally used for construction of the peak type approximation.
 All quantities at response functions are estimated in peak, and
subsequent integration of response functions leads to results in terms
of suppression
factors $S_{E,M,EM}$ (or of sum rules for electron-nucleus scattering
\cite{LLS}). Here we give explicit formulae for $^3$He target:
\begin {equation}
\begin {array}{l}
\displaystyle
\Im ^{q}_{1}= \eta (\mu_n^2+2\mu_p^2)S_M,\\[0.5cm]
\displaystyle
\Im ^{q}_{2}=\od{\eta (\mu_n^2+2\mu_p^2)S_M+(e_n^2+2e_p^2)S_E}
{1+\eta },\\[0.5cm]
\displaystyle
\Im ^{q}_{3}= \od{P_N}2(P_ne_n\mu_n+2P_pe_p\mu_p)S_{EM},
\\[0.5cm]
\displaystyle
\Im ^{q}_{4}=  {P_N\over 4}{(P_ne_n\mu_n+2P_pe_p\mu_p)S_{EM}-
(P_n\mu_n^2+2P_p\mu_p^2)S_M \over 1+\eta }
,
\end {array}
\label {qf}
\end {equation}
where $\eta$ is $\eta_A$ for nucleon.
$P_p$ and $P_n$ are effective proton and neutron polarization in $^3$He
and
$e_{p,n}$, $\mu_{p,n}$ are electric and magnetic formfactors of proton
and neutron.

\subsection{Electroweak and semi-inclusive structure
functions}
\label{appA2}

Let us define the fermion vertexes and boson propagators.
We introduce
\begin{equation}
\begin{array}{c}
\displaystyle
 v^{\gamma }_f=-e_f,\;a^{\gamma }_f=0,\;
   \\[0.5cm]
\displaystyle
 v^Z_f={I^3_f-2s^2_We_f\over 2s_Wc_W} ,\;
 a^Z_f={I^3_f\over 2s_Wc_W} ,\;
   \\[0.5cm]
\displaystyle
  v^W_f=a^W_f={1\over 2\sqrt {2} s_W},
\end{array}
\end{equation}
where $e_f$ and $I^3_f$ --- electric charge and the third
component of the fermions weak isospin,
$s_W$ and $c_W$ --- Weinberg angle sine and cosine respectively.
Hence, the fermion vertexes takes the form
\begin {equation}
-ie\ga_{\mu}(v^i_f-a^i_f\gamma _5).
\end {equation}
The next couple constants combinations are contained in
observables
\begin{equation}
\begin{array}{ll}
\displaystyle
\lambda^{f ij}_V=2(v^i_fv^j_f+a^i_fa^j_f),\qquad
\lambda^{f ij}_A=2c_f(v^i_fa^j_f+a^i_fv^j_f)\; ,
\\[0.5cm] \displaystyle
R^{ij}_+=\lambda^{e ij}_V-P_L\lambda^{e ij}_A,\qquad\;\;\;
R^{ij}_-=\lambda^{e ij}_A-P_L\lambda^{e ij}_V\; ,
\\[0.5cm] \displaystyle
F^{ij}_+(x,Q^2)=
\chi^{n_Z}
\sum_q \left [\lambda^{qij}_Vxf^{(+)}_q(x,Q^2)
+P_N\lambda^{qij}_Axf^{(-)}_q(x,Q^2)\right ],
\di5
F^{ij}_-(x,Q^2)=
\chi^{n_Z}
\sum_q \left [\lambda^{qij}_Axf^{(+)}_q(x,Q^2)
+P_N\lambda^{qij}_Vxf^{(-)}_q(x,Q^2)\right ],
\end{array}
\label{f1g1ew}
\end{equation}
where
$c_f=1(-1)$ for fermions (antifermions), $\chi=Q^2/(Q^2+M_z^2)$
and
$n_z=(0,1,2)$ for $ij=\gamma \gamma $, $\gamma Z$ or $Z\gamma $,
$ZZ$.
In QCD-improved model the
parton distributions $f^{(\pm)}_q(x,Q^2)$ depend on  $Q^2$:
\begin{equation}
f^{(\pm)}_q(x,Q^2)=
f^{\uparrow \uparrow }_q(x,Q^2)
\pm f^{\uparrow \downarrow }_q(x,Q^2)
,\end{equation}
where  $f^{\uparrow \downarrow
}_q(x,Q^2)$ and $f^{\uparrow \uparrow }_q(x,Q^2)$
--- densities of type $q$ partons with helicities,
(anti)parallel to nucleon helicity respectively.

The quantities
\begin{equation}
\Sigma ^{+(-)}(x,z) = \sum^{}_{q} e^{2}_{q} [f^{+}_{q}(x) \pm
f^{-}_{q}(x)] D^{H}_{q}(z)
\end{equation}
is the ordinary for QPM combination of the distribution  functions
$f^{+(-)}_{q}(x)$   for   the   quark   of  flower  $q$  polarized
(anti)parallel   to   the   nucleon   polarization,   and  of  the
fragmentation functions  $D^{H}_{q}(z)$ of  the quark  q into  the
hadron $H$, $e_{q}$ being the quark charge in units of  elementary
charge.

\subsection{Fits and models for structure
functions}
\label{appA3}

    RC calculation  requires fits or models for SF, elastic
formfactors, quasielastic suppression factors, fragmentation
functions and partonic densities to be known in a whole region of
varying of integration variables.

POLRAD 2.0 gives the opportunity to choose between three models of
spin-average and spin-dependent proton and neutron (deuteron) SF
$F_2^{p,d}(x, Q^2)$ and
$g_1^{p,d}(x, Q^2)$.

For the explanation of the first model for unpolarized SF see
fig.\ref{SFF2}. In the small
$Q^2$ region parametrizations of
the ref.\cite {Bra}(in resonance region) and ref.\cite {Ste}  are
used, and for all the rest kinematics  15-parameter
NA-47 fit
\cite {Mil} is adopted.
The advantage of this model is the implementation of the modern
experimental data in small $x$ regions. For
the calculation of $R(x,Q^{2})$ and $F^{d}_{2}/F^{p}_{2}$
the fits shown on the corresponding grafs are used.

The second model is based on the fit of \cite {D8} with the modern
parameters and has the simple analytical form, same for the all
kinematical region. However, it does not give good description of
the modern data when
$x>0.01$. In this model for $R$ we adopt
the Whitlow fit $R^{1990}$ \cite {Whi}.

For both models
$g_1(x,Q^2)=F_1(x,Q^2) A_1(x,Q^2)$ and $A_1(x,Q^2)$ could be
obtained either from the iteration procedure data or from the
asymmetry fit (\ref{0014},\ref{00d7}).

In the third model partonic distrbutions (with \cite{GRV,GRSV96}
or
without \cite{Sch} taking into account $Q^2$ dependence) are used
for the construction of
$F_2^{p,d}(x, Q^2)$ and $g_1^{p,d}(x, Q^2)$. This fit is
usually used in experiments at collider.

$F^{A}_{2}(x,Q^{2})$  for other  nuclei  is  calculated in
accordance with ref. \cite{Smi}
and $R$ is considered  to
be A-independent (see review \cite {Arn}, for example).
Convolution expressions
for   $g^d_{1}$, $g^{^3{\rm He}}_{1}$  are obtained via
$g^{p}_{1}(x)$ and
$g^{n}_{1}(x)$  \cite{BV}.

For $g_2$ one can choose two possible variants:
simple partonic approximation $g_2=0$ and the Wandzura-Wilczek
relation \cite{WW}
\begin{equation}
g_2(x,Q^2)=-g_1(x,Q^2)+ \int \limits^{1}_{x} \frac{dz}{z}
g_1(z,Q^2).
\end{equation}

Quadruple SF $b_{1},b_{2}$ should be  taken  into  account  for
deuterons  as  spin-1  particles.  They   are   related   by
Callan-Gross equality
\begin {equation}
b_{2}(x) = 2xb_{1}(x)
\end {equation}
\noindent and conform to a sum rule obtained  in  ref.\cite {CK}.  Model  of
ref.\cite {KH} is used for them.

Deuteron formfactors are calculated in accordance with the model
of ref.\cite{Kob}, which provides the right asymptotic behavior.
For $^{3}$He
form-factors we use fit from ref.\cite{ffhe}.
Nucleon  formfactors are taken  from
ref.\cite {Bil}.
Charge formfactor for scalar nuclei can be found in ref.\cite {Akh}
(see sect. 3.8).

The suppression factors for QRT  for  DIS  on  deuteron
target are calculated as the same in ref.\cite {Ste}.
For other
nuclei we used the Fermi gas model \cite {De,Mon}.

There are three possible models for the fragmentation functions
\cite{AUB, CMPB, ffarn} in SIDIS calculation.

\subsection{Parameterization of neutron spin asymmetry}
\label{appA4}
The following function taken from Schaefer's parameterization \cite{Sch}
is used for fitting of the neutron spin asymmetry:
\beq
A_1^n(x)=\od1{a_0+3a_1}\left(a_0f_d^0+\od{a_1}{9}\bigl(
-16f_u^1+8f_u^0-2f_d^1+f_d^0\bigr)\right),
\eeq{a1}
where
\barr{l}
a_0={2x^{\al_u}(1-x)^{\be_u}\over B (\al_u,\be_u+1)}
-{x^{\al_d}(1-x)^{\be_d}\over 2B (\al_d,\be_d+1)}
,
\di3
a_1=\od32{x^{\al_d}(1-x)^{\be_d}\over B (\al_d,\be_d+1)}
\earr{a2}
and
\barr{ll}
f_u^0={1\over 1+a_{u0}x^{\al_u}(1-x)^2}
,\dis
f_u^1={1\over 1+a_{u0}a_{10}x^{\al_u}(1-x)^2}
,\di3
f_d^0={1\over 1+a_7a_{u0}x^{\al_d}(1-x)^2}
,\dis
f_d^1={1\over 1+a_7a_{u0}a_{10}x^{\al_d}(1-x)^2}
.\earr{a3}
Only parameters $a_7$, $a_{u0}$ and $a_{10}$ are fitted,
and $\al_u$, $\al_d$, $\be_u$ and $\be_d$ are considered to be constant
\beq
\al_u=0.588, \quad \al_d=1.03, \quad
\be_u=2.69, \quad \be_d=6.89.
\eeq{a35}

\section{Quantities $\Theta_{ij}(\tau)$}
\label{appB}
\setcounter{equation}{0}

In this Appendix we give  the  explicit  form  for  the
functions $\theta _{ij}(\tau)$, which are
 contained in the final  formulae
for the radiative tails. $i$ runs from 1 to  8.  This  fact
corresponds to the contributions of eight SF  or  formfactor
combinations, and $j$  runs from 1 to $k_{i}$  which  are
defined in \r{kili}. The function $\theta _{ij}(\tau)$ can be
found as a sum over $k$
\beq
\theta _{ij}(\tau)=\sum_k a_{ik}T_{ijk}(\tau),
\eeq{th01}
which is calculated from
$max(1,j+l_i-k_i)$ to $min({j,l_i})$, where
\beq
k_{i} = (3,3,4,5,5,5,3,4), \qquad
l_{i} = (1,1,1,2,3,3,1,2),
\eeq{kili}
and
\beq
a_{ik}
=\left\{
   \begin{array}{ccl}
   \displaystyle
    1\;\;
   &\displaystyle
    {\rm for}\; k=1,
   &\displaystyle
     i=1,2,3,7;
    \\[3mm]
   \displaystyle
     \{\eta q,-1\}\;\;
   &\displaystyle
     {\rm for}\; k=\{1,2\},
   &\displaystyle
     i=4,8;
    \\[3mm]
   \displaystyle
     \{Q^2-3(\eta q)^2,6\eta q,-3\}\;\;
   &\displaystyle
     {\rm for}\; k=\{1,2,3\},
   &\displaystyle
     i=5,6.
   \end{array}
\right.
\eeq{th02}

The quantities $T_{ijk}(\tau)$ for $k=1$ take the form
\begin{eqnarray}
T _{111}(\tau)&=& 4(Q^2-2m^{2})F_{IR},
\nonumber\\[0.3cm]
T _{121}(\tau)&=& 4\tau F_{IR},
\nonumber\\[0.3cm]
T _{131}(\tau)&=& -4F - 2\tau^{2}F_{d},
\nonumber\\[0.3cm]
T _{211}(\tau)&=& {2(SX - M^{2}Q^2)F_{IR}/M^{2}},
\nonumber\\[0.3cm]
T _{221}(\tau)&=&
\pmatrix{2m^{2}S_{p}F_{2-}+S_{p}S_{x}F_{1+}+2(S_{x}-2M^{2}\tau)F_{IR}-\tau S
 ^{2}_{p}F_{d}}/2M^{2},
\nonumber\\[0.3cm]
T _{231}(\tau)&=&
\pmatrix{4M^{2}F+(4m^{2}+2M^{2}\tau^{2}-S_{x}\tau)F_{d}-S_{p}F_{1+}}
/2M^{2},
\nonumber\\[0.3cm]
T _{311}(\tau)&=&{-8P_{L}m\over M}(\eta q\; k_{2}\xi
- Q^2\; \xi \eta )F_{IR},
\nonumber\\[0.3cm]
T _{321}(\tau)&=&{2P_{L}m\over M}\pmatrix{\eta {\cal K}
(4m^{2}F^{\xi }_{d}
 -4m^{2}F^{\xi }_{2+}+2F^{\xi }_{IR}-Q^2F^{\xi
 }_{2-}+Q^2_{m}F_{2+}^{\xi }}
+\nonumber\\[0.3cm]
&&+ k_{2}\xi  (-8m^{2}F^{\eta }_{d}
+ 4m^{2}F^{\eta }_{2+}+ 2\;\eta q \;\tau F_{d})-
 4\;\eta q\; F^{\xi }_{IR}+ 4\;\xi \eta\;  \tau F_{IR}),
\nonumber\\[0.3cm]
T _{331}(\tau)&=&{2P_{L}m\over M}(\eta {\cal
 K}\;\tau (F^{\xi }_{2+}-F^{\xi }_{2-}-2F^{\xi
 }_{d})
 +\nonumber\\[0.3cm]
&&2\;k_{2}\xi\;
  \tau F^{\eta }_{d} + 4m^{2}F^{\xi \eta }_{d}
+ 6F^{\xi \eta }_{IR} + Q^2F^{\xi \eta }_{2-} - Q^2_{m}F^{\xi \eta
 }_{2+}),
\nonumber\\[0.3cm]
T _{341}(\tau)&=&{-2P_{L}m\tau \over M}\pmatrix{2F^{\xi \eta
 }_{d}+F^{\xi \eta }_{2+}-F^{\xi \eta }_{2-}},
\nonumber\\[0.3cm]
T_{411}(\tau)&=&{4mP_{L}\over M^{2}}\pmatrix{S_{x}\;\xi k_{2}
-2\;\xi
 p\; Q^2}F_{IR},
\nonumber\\[0.3cm]
T_{421}(\tau)&=&{mP_{L}\over M^{2}}(4m^{2}
 (2\;k_{2}\xi\; F_{d}-k_{2}\xi\; F_{2+}-S_{p}F^{\xi
 }_{d}+S_{p}F^{\xi }_{2+})- 2\;k_{2}\xi\;
\tau S_{x}F_{d}
\nonumber\\[0.3cm]
&&- 8\;\xi p\; \tau F_{IR}+2\pmatrix{2S_{x}-S_{p}}F^{\xi }_{IR}+
 S_{p}\pmatrix{Q^2F^{\xi }_{2-}-Q^2_{m}F^{\xi }_{2+}}),
\nonumber\\[0.3cm]
T_{431}(\tau)&=&{-mP_{L}\over M^{2}} (
 2F^{\xi }_{d} \pmatrix{2m^{2}-\tau S_{p}} + 2\;k_{2}\xi\;
  \tau F_{d}+ 6F^{\xi}_{IR}
\\[0.3cm]
&& + \pmatrix{Q^2-\tau S_{p}}F^{\xi }_{2-}
- \pmatrix{Q^2_{m}-\tau S_{p}}F^{\xi }_{2+}),
\nonumber\\[0.3cm]
T_{441}(\tau)&=&{mP_{L}\tau\over M^{2}}\pmatrix{2F^{\xi
}_{d}-F^{\xi
 }_{2-}+F^{\xi }_{2+}},
\nonumber\\[0.3cm]
T _{711}(\tau)&=& -2\pmatrix{Q^2+4m^{2}+12\;\eta k_{1}\;\eta
k_{2}}F_{IR},
\nonumber\\[0.3cm]
T _{721}(\tau)&=& -2\pmatrix{3\;\eta {\cal K}
(2m^{2}F_{2-}^{\eta }-\eta {\cal K}\;\tau F_{d}+\eta q\;F
 _{1+})+6\;\eta q\;F^{\eta }_{IR}+\tau F_{IR}},
\nonumber\\[0.3cm]
T _{731}(\tau)&=& 2F + \tau ^{2}F_{d}+ 6\pmatrix{\eta {\cal
K}\;F^{\eta }
_{1+}+\eta q\;\tau F^{\eta }_{d}-4m^{2}F^{\eta \eta }_{d}},
\nonumber\\[0.3cm]
T_{811}(\tau)&=&{ -6\over M} \pmatrix{S\;\eta k_{2}+X\;\eta
 k_{1}}F_{IR},
\nonumber\\[0.3cm]
T_{821}(\tau)&=&{- 3\over M}(\eta
 {\cal K} \pmatrix{m^{2}F_{2-}-\tau S_{p}F_{d}}+ \eta q\; F_{IR}
 + m^{2}S_{p}F^{\eta }_{2-}
\nonumber\\[0.3cm]
&&+ \pmatrix{S\;\eta k_{1}+X\;\eta k_{2}}F_{1+}+ S_{x}F^{\eta
 }_{IR}),
\nonumber\\[0.3cm]
T_{831}(\tau)&=&{-3\over 2M}\pmatrix{8m^{2}F^{\eta }_{d}-\eta
 {\cal K}\;F_{1+}-\eta q\;\tau F_{d}-S_{x}\tau
 F^{\eta }_{d}-S_{p}F^{\eta }_{1+}}.
\nonumber
\label{th15}
\end{eqnarray}

\noindent For $i =5$ and $i =6$ we have
\beq
T_{\{5,6\}j1}(\ta )=T_{\{1,2\}j1}(\ta ).
\eeq{th04}
The quantities $T_{ijk}(\tau)$ for $k=2,3$ are calculated as
\beq
T_{ijk}(\ta )=T_{ij-1k-1}(\ta )\left\{
F_{all} \rightarrow F_{all}^{\eta},\;
F_{all}^{\xi} \rightarrow F_{all}^{\xi\eta},\;
F_{all}^{\eta} \rightarrow F_{all}^{\eta\eta}
\right\}+
q_{ik}T_{ij-1k-1}(\ta ).
\eeq{th06}
The second term appears only for $i=5,6$ and $k=2$:
\beq
q_{ik}=
\delta_{k2}
(\delta_{i5}+
\delta_{i6}) \od{\ta}M
.
\eeq{th08}
The substitution in the first term of \r{th06} has to be applied
for
all $F$ contained in $T_{ij-1k-1}(\ta )$. The quantities $F$ with
an upper index are obtained in terms of $F$ without the index:
\begin{eqnarray}
2F_{2+}^{\{\xi,\eta\}}&=&
(2F_{1+}+\ta F_{2-})s_{\{\xi,\eta\}}+F_{2+}r_{\{\xi,\eta\}},
\nonumber\\[0.3cm]
2F_{2-}^{\{\xi,\eta\}}&=&(2
F_{d}+F_{2+})\ta s_{\{\xi,\eta\}}+F_{2-}r_{\{\xi,\eta\}},
\nonumber\\[0.3cm]
2F_{d}^{\{\xi,\eta\}}&=&F_{1+}s_{\{\xi,\eta\}}+F_{d}r_{\{\xi,\eta\}},
\nonumber\\[0.3cm]
4F_{2+}^{\{\xi,\eta\}\eta}&=&
(2F_{1+}+\ta F_{2-})
(r_{\eta}s_{\{\xi,\eta\}}+s_{\eta}r_{\{\xi,\eta\}})
+F_{2+}
(r_{\eta}r_{\{\xi,\eta\}}+\ta^2 s_{\eta}s_{\{\xi,\eta\}})
\nonumber\\[0.3cm]
&&+4(2F+F_d\ta^2)s_{\eta}s_{\{\xi,\eta\}},
\nonumber\\[0.3cm]
4F_{2-}^{\{\xi,\eta\}\eta}&=&
(2F_{d}+ F_{2+})
(r_{\eta}s_{\{\xi,\eta\}}+s_{\eta}r_{\{\xi,\eta\}})
+F_{2-}
(r_{\eta}r_{\{\xi,\eta\}}+\ta^2 s_{\eta}s_{\{\xi,\eta\}})
\nonumber\\[0.3cm]
&&+4\ta F_{1+}s_{\eta}s_{\{\xi,\eta\}},
\\[0.3cm]
4F_{d}^{\{\xi,\eta\}\eta}&=&
F_{1+}
(r_{\eta}s_{\{\xi,\eta\}}+s_{\eta}r_{\{\xi,\eta\}})
+F_{d}
(r_{\eta}r_{\{\xi,\eta\}}+\ta^2 s_{\eta}s_{\{\xi,\eta\}})
+4 Fs_{\eta}s_{\{\xi,\eta\}},
\nonumber\\[0.3cm]
2F_{1+}^{\eta}&=&(4F+\ta^2F_d)s_{\eta}+F_{1+}r_{\eta},
\nonumber\\[0.3cm]
4F_{1+}^{\eta\eta}&=&2(4F+\ta^2F_d)r_{\eta}s_{\eta}
+F_{1+}(r_{\eta}^2+\ta^2s_{\eta}^2)
+4(2F_i-\ta F)s_{\eta}^2),
\nonumber\\[0.3cm]
2F^{\eta}&=&F(r_{\eta}-\ta s_{\eta})+2F_{i}s_{\eta},
\nonumber
\\[0.3cm]
4F^{\eta\eta}&=&F(r_{\eta}-\ta s_{\eta})^2
+4F_{i}(r_{\eta}-\ta s_{\eta})
+4F_{ii}s_{\eta}^2.
\nonumber
\end{eqnarray}
The quantities
\beq
s_{\{\xi,\eta\}}=a_{\{\xi,\eta\}}+b_{\{\xi,\eta\}},
r_{\{\xi,\eta\}}=\ta (a_{\{\xi,\eta\}}-b_{\{\xi,\eta\}})
+2c_{\{\xi,\eta\}}
\eeq{th10}
are combinations of coefficients of polarization vectors
$\xi$ and
$\eta$ expansion over basis (see sect.\ref{2.01})
\beq
\xi,\eta=2(
a_{\{\xi,\eta\}} k_1 +
b_{\{\xi,\eta\}} k_2 +
c_{\{\xi,\eta\}} p).
\eeq{th11}
We note that the scalar products from \r{th15} and \r{bt} are also
calculated in terms of the coefficients
\barr{c}
\eta q =-Q^2(a_{\eta}-b_{\eta})+S_xc_{\eta},   \quad
\eta {\cal K} =(Q^2+4m^2)(a_{\eta}+b_{\eta})+S_pc_{\eta},   \di5
2\eta k_1=\eta {\cal K}+\eta q,   \quad
2\eta k_2=\eta {\cal K}-\eta q,    \di5
k_2\xi =Q^2_ma_{\xi}+2m^2b_{\xi}+Xc_{\xi},   \quad
\xi p =Sa_{\xi}+Xb_{\xi}+2M^2c_{\xi},   \di5
\ot12 \xi\eta=
2m^2(a_{\xi}a_{\eta}+b_{\xi}b_{\eta})
+2M^2c_{\xi}c_{\eta}
+Q_m^2(a_{\xi}b_{\eta}+b_{\xi}a_{\eta})
+S(a_{\xi}c_{\eta}+c_{\xi}a_{\eta})
+X(b_{\xi}c_{\eta}+c_{\xi}b_{\eta}).
\earr{th16}

The following equalities define the functions $F$:
\barr{ll}
F =   \lambda ^{-1/2}_{Q},
\dis
F_{IR} = m^{2}F_{2+}- Q^2_mF_{d},
\di5
F_{d} =   \tau ^{-1} (C^{-1/2}_{2}(\tau )-C^{-1/2}_{1}(\tau ))
\dis
F_{1+} =  C^{-1/2}_{2}(\tau ) + C^{-1/2}_{1}(\tau ),
\di5
F_{2+} = B_2(\tau ) C^{-3/2}_2(\tau ) - B_1(\tau ) C^{-3/2}_1(\tau ),
\dis
F_{2-} = B_2(\tau ) C^{-3/2}_2(\tau ) + B_1(\tau ) C^{-3/2}_1(\tau ),
\di5
F_{i} = - \lambda ^{-3/2}_{Q} B_{1}(\tau ),
\dis
F_{ii} =
{1\over 2}\lambda ^{-5/2}_Q
(3B_{1}^2(\tau )-\lambda _{Q}C_{1}(\tau )),
\earr{th20}
 where
\begin{equation}
\begin{array}{l}
\displaystyle
B_{1,2}(\tau )= - {1\over 2} \pmatrix{\lambda _{Q}\tau \pm
 S_{p}(S_{x}\tau +2Q^2)},
 \\[0.5cm]\displaystyle
C_{1}(\tau ) =(S\tau + Q^2)^{2}+ 4m^{2}(Q^2 + \tau S_x - \tau^{2}M^{2}),
 \\[0.5cm]\displaystyle
C_{2}(\tau ) =(X\tau - Q^2)^{2}+ 4m^{2}(Q^2 + \tau S_x - \tau^{2}M^{2}).
\end {array}
\end {equation}
We note that $F_d$ has a uncertainty like $0/0$ for $\ta=0$
(inside of integration region). It leads to difficulties for
numerical integration, so the another form is used also
\beq
F_d={S_p(\ta S_x+2Q^2)\over {C_1^{1/2}(\tau )}{C_2^{1/2}(\tau )}
({C_1^{1/2}(\tau )}+{C^{1/2}_2(\tau )})}.
\eeq{th22}

\section{Quantities ${\cal R}_{1,2}^{u,p}$}
\label{appC}
\setcounter{equation}{0}

The functions ${\cal R}_{1,2}^{u,p}$ contributed to \r{qq20} are
listed here:
\def\lny{L_y}
\def\lne{L_1}
\def\dne{D_1}
\def\lnye{L_{y1}}
\def\dnye{D_{y1}}
\def\sqq{(2f(\eta)-1)}

\begin{eqnarray}
{\cal R}_{1}^{u}&=&-2Y_+{\tilde X}{\tilde L}
+2Y_+((\od x 3-\od3x)f(\eta)-\od2x+\od{14}{3x^2})
-\od 23(2xf(\eta)+\od 1{x^2}-\od 6x)
\nonumber\\[0.3cm] &&
+ (\od x3+\od {4x}3f(\eta)-3){y^2 \over 2y_1\eta}
  -\od2{x^2}(1-x)(\od 1{y_1}+Y_+)\lne
  + \od1{2\eta}(2y_1+Y_+)\lnye
\nonumber\\[0.3cm] &&
  + \od1{2\eta}(\od2{y_1}+Y_+)\lne
  - \od2{\eta}Y_+\lny
  - \od2{x^2}(1-x)(y_1+Y_+)\lny
        + \od{2y_1}{x^2y}(  xy + y_1)\dne
\nonumber\\[0.3cm] &&
         - \od{y}{2\eta}(\dne+\dnye)
  + \od{2}{yx^2}(1 - xy)\dnye,
\nonumber\\[0.3cm]
{\cal R}_{2}^{u}&=&-2Y_+{\tilde L}
+ \od{y^2}{y_1}(2xf(\eta)-3)
+ (\od 2{y_1}+Y_+)\lne
+ (2y_1+Y_+)\lnye
- 4Y_+\lny
\nonumber\\[0.3cm]
&&- y(\dne +\dnye),
\nonumber\\[0.3cm]
{\cal R}_{1}^{p}&=&2Y_-x{\tilde X}{\tilde L}
-2Y_-(1-\od 3x+ 2f(\eta))+
 2(Y_-(1-\od 1x)-\od1{xy_1})\lne+
\nonumber\\[0.3cm] &&
 2(Y_-(1-\od 1x)+\od{y_1}x)\lnye
 +\od x{2\eta}(Y_-+\od{2y}{y_1})\lne
 -\od x{\eta}Y_-Y_+\lne-\od {xy}{2\eta}(2Y_+-\od y{y_1})\lnye
\nonumber\\[0.3cm] &&
+ (2y + 2\od{y_1}{x}-\od{y^2x}{2y_1\eta})\dne
+ (2\od 1{x}-2y-\od{y^2x}{2y_1\eta})\dnye,
\nonumber\\[0.3cm]
{\cal R}_{2}^{p}&=&2Y_-{\tilde L}
+Y_-(\lne -1+\lnye)+y(\dne-\dnye),
\end{eqnarray}
where
\barr{c}
\lny={\ln(xf(\eta))},   \quad
\lne={\ln(y_1+xyf(\eta))},   \quad
\lnye={\ln({1-xyf(\eta)\over y_1})},  \di5
\dne={1\over y_1+xyf(\eta)}, \quad
\dnye={-1\over 1-xyf(\eta)},  \di5
{\tilde
L}=\ln\od{(1-x)^2}{(1-xf(\eta))^2(1-xy)(y_1+xy)}+\lne+\lnye
\earr{qq21}
and $2f(\eta)=1+\sqrt{1/\eta+1}$.

\section{Test Run Output}
\label{appD}

\subsection{Example 1. Output file ITDAT3.DAT}
\label{appD1}
\begin{verbatim}
*************** Helium-3 data **********************
   .010   .830  -.03500  -.04566  -.03500   .02000
   .020   .790  -.05500  -.06745  -.05500   .02100
   .035   .740  -.07500  -.08794  -.07500   .02300
   .050   .660  -.10000  -.11211  -.10000   .02700
   .070   .620  -.08000  -.09195  -.08000   .02800
   .095   .570  -.12800  -.13951  -.12800   .03000
   .115   .530  -.14000  -.15128  -.14000   .03200
   .140   .480  -.16000  -.17111  -.16000   .03400
   .180   .450  -.12500  -.13620  -.12500   .03500
   .230   .425  -.17000  -.18141  -.17000   .03600
   .285   .400  -.14500  -.15666  -.14500   .03800
   .350   .360  -.12500  -.13701  -.12500   .04000
   .430   .330  -.08000  -.09231  -.08000   .04200
   .530   .300   .00800  -.00438   .00800   .04400
   .640   .270   .09000   .07667   .09000   .04600
   .745   .240   .22000   .20203   .22000   .04800
   .850   .220   .48800   .45913   .48800   .05000
Next two lines are additional data for fit constructing.
* 1.000  0.000  1.00000  1.00000  1.00000   .00000
*  .000   .000   .00000   .00000   .00000   .00000
Am - measured spin asymmetry,
Al - last step spin asymmetry,
Ap - previous step spin asymmetry,
Err- experimental error.
\end{verbatim}

\subsection{Example 2. Output file ASM.DAT}
\label{appD2}

\begin{verbatim}
 program polrad20 version from 10.04.1997

 the file gives born asymmetry, observed asymmetry
 and radiative correstion

 the following switches are active
   polrad strffun integrat polrad_add fits2 alpha2ll exact elect long
   kin_net targ_h f2g1grsv96 g1asym g2_eq_0 pol_asym onlyin
   electroweak

 leptons are electrons
 target is proton
 target is longitudinally polarized
 bmom =  27.5
 tmom = 830.0
    pl = 1.00    pn = 1.00   qn = 0.00

 a is in %

      x      w2       q2  a(born)  a(obs)  del(%)
   0.001   913.0     0.9   0.378   0.334  -0.043
   0.001  1369.0     1.4   0.378   0.337  -0.040
   0.001  1825.1     1.8   0.378   0.339  -0.038
   0.001  2737.1     2.7   0.378   0.342  -0.035
   0.001  3649.2     3.7   0.378   0.345  -0.032
   0.001  4561.3     4.6   0.378   0.347  -0.030
   0.001  5473.4     5.5   0.378   0.349  -0.029
   0.001  6385.5     6.4   0.378   0.349  -0.028
   0.001  7297.6     7.3   0.378   0.350  -0.027
   0.001  8209.7     8.2   0.378   0.351  -0.027
   0.001  9121.8     9.1   0.378   0.351  -0.027
   0.001 11402.0    11.4   0.378   0.352  -0.026
   0.001 13682.2    13.7   0.378   0.352  -0.025
   0.001 15962.4    16.0   0.378   0.353  -0.025
   0.001 18242.6    18.3   0.378   0.353  -0.024
   0.001 27363.5    27.4   0.378   0.355  -0.023
   0.001 36484.4    36.5   0.378   0.355  -0.022
   0.001 45605.2    45.7   0.378   0.355  -0.022
   0.001 54726.1    54.8   0.378   0.356  -0.021
   0.001 59286.6    59.3   0.378   0.355  -0.023
   0.001 63847.0    63.9   0.378   0.353  -0.025
   0.001 68407.4    68.5   0.378   0.350  -0.027
   0.001 72967.9    73.0   0.378   0.346  -0.032
   0.001 75248.1    75.3   0.378   0.343  -0.035
   0.001 77528.3    77.6   0.378   0.341  -0.037
   0.001 79808.5    79.9   0.378   0.336  -0.041
   0.001 82088.7    82.2   0.378   0.336  -0.041
   0.001 83000.8    83.1   0.378   0.324  -0.054
   0.001 83912.9    84.0   0.378   0.323  -0.055
   0.001 84825.0    84.9   0.378   0.320  -0.058
   0.001 85737.1    85.8   0.378   0.310  -0.067
   0.001 86649.2    86.7   0.378   0.298  -0.079
   0.001 87561.3    87.6   0.378   0.291  -0.087
   0.001 88473.3    88.6   0.378   0.278  -0.099
   0.001 89385.4    89.5   0.378   0.261  -0.116
   0.001 90297.5    90.4   0.378   0.235  -0.142
\end{verbatim}

\subsection{Example 3. Output file ASM.DAT (with CUTS)}
\label{appD3a}

\begin{verbatim}
 program polrad20 version from 10.04.1997

 the file gives born asymmetry, observed asymmetry
 and radiative correstion

 the following switches are active
   sirad strffun integrat elect long kin_net targ_h f2g1grsv96 g1asym
   qdstr_gu g2_eq_0 ffrg_aub pol_asym onlyin outfun_a intdz cuts eweak
   pi_diff

 leptons are electrons
 target is proton
 target is longitudinally polarized
 bmom =  27.5
 tmom =    .0
    pl = 1.00    pn = 1.00   qn =  .00

 a is in %
      x      y    q**2      A1     rc A1     born     meas
    .115   .421  2.490   58.912   -2.828   30.237   29.382
    .183   .421  3.980   61.646   -1.666   31.289   30.768
    .252   .421  5.470   63.000    -.986   31.780   31.467
    .320   .421  6.960   63.683    -.574   32.015   31.831
    .389   .421  8.449   63.999    -.322   32.116   32.013
    .457   .421  9.939   64.005    -.198   32.103   32.040
    .526   .421 11.429   63.979    -.171   32.106   32.051
    .594   .421 12.919   63.938    -.298   32.150   32.054
\end{verbatim}

\subsection{Example 3. Output file ASM.DAT (without CUTS)}
\label{appD3b}

\begin{verbatim}
 program polrad20 version from 10.04.1997

 the file gives born asymmetry, observed asymmetry
 and radiative correstion

 the following switches are active
   sirad strffun integrat elect long kin_net targ_h f2g1grsv96 g1asym
   qdstr_gu g2_eq_0 ffrg_aub pol_asym onlyin outfun_a intdz eweak
   pi_diff

 leptons are electrons
 target is proton
 target is longitudinally polarized
 bmom =  27.5
 tmom =    .0
    pl = 1.00    pn = 1.00   qn =  .00

 a is in %
      x      y    q**2      A1     rc A1     born     meas
    .115   .421  2.490   57.759   -4.730   30.237   28.807
    .183   .421  3.980   60.751   -3.093   31.289   30.321
    .252   .421  5.470   62.364   -1.985   31.780   31.149
    .320   .421  6.960   63.256   -1.241   32.015   31.618
    .389   .421  8.449   63.742    -.722   32.116   31.885
    .457   .421  9.939   63.908    -.349   32.103   31.991
    .526   .421 11.429   64.045    -.069   32.106   32.084
    .594   .421 12.919   64.212     .129   32.150   32.191
    .663   .421 14.409   64.189     .303   32.106   32.204
    .731   .421 15.899   64.364     .481   32.160   32.315
\end{verbatim}

\newpage
FIGURE CAPTIONS
\begin{enumerate}
\item
The double differential cross section for inclusive
lepton-nuclei scattering sketched at a certain value of $Q^2$ as
function of $\protect\nu$ (arbitrary scale). The three basic channels
are the elastic (I), the quasielastic (II), and the inelastic (III)
one.
\item
The limits of integration for variables
a) $R$ and $\protect\ta$ \protect\r{eq25};
b) $\protect\xi$ and $t$ (\protect\ref{0009b}-\protect\ref{LLL});
b) $t_1$ and $t_2$ \protect\r{sigFR}.
\item
Scheme of program POLRAD. The subroutine APPTAI, AL2LL
and TARGWS are presented on a separate figures.
\item
Scheme of program SIRAD.
\item
Scheme of program AL2LL.
\item
Scheme of program APPTAI.
\item
Scheme of program TARGWS.
\item
The results of iteration procedure for spin asymmetry (see
sect.\ref{exampleit}) along with the fit constructed.
\item
The unpolarization a) and polarization b) radiative correction
factors defined by eqn.\protect\r{RCF}. The curves 1,2 and 3
correspond to different values of $x=\;0.001$, $0.01$, $0.1$
respectively.
\item
The born (dash) and observed (solid) SIDIS asymmetries (a) and
relative correction with (dash) and without (solid) kinematical
cuts apllied (b).
\item
The spin-independent proton and neutron SF. The
citations denote that fit from the cited references is used in the
range. The fits are extrapolated into hatched region. The procedure
of joining together of two-dimensional surfaces gives continuous
fits in the whole kinematical region.
\end{enumerate}

\newpage

\input{epsf}

\hspace*{2.5cm}
\newbox\plrbox
\newdimen\plrwd
\font\plra=plra at 72.27truept
\setbox\plrbox=\vbox{\hbox{%
\plra\char0\char1\char2}}
\plrwd=\wd\plrbox
\setbox\plrbox=\hbox{\vbox{\hsize=\plrwd
\parskip=0pt\offinterlineskip\parindent0pt
\hbox{\plra\char0\char1\char2}
\hbox{\plra\char3\char4\char5}
\hbox{\plra\char6\char7\char8}}}
\ifx\parbox\undefined
    \def\setplr{\box\plrbox}
\else
    \def\setplr{\parbox{\wd\plrbox}{\box\plrbox}}
\fi

\setplr \\

\vspace*{0.3cm}

\begin{figure}[h]
\caption{
}
\label{chan}
\end{figure}

\newpage
\begin{figure}[p]
\vspace{5cm}
\begin{tabular}{ccc}
\begin{picture}(150,100)
\put(10,10){\vector(2,0){80.}}
\put(30,10){\vector(0,2){80.}}
\put(20,10){\line(0,2){50.}}
\put(70,10){\line(0,2){30.}}
\put(70,65){\oval(100.,50.)[lb]}
\put(20,80){\makebox(0,0){$R$}}
\put(90,0){\makebox(0,0){$\ta$}}
\put(50,-10){\makebox(0,0){\small a)}}
\end{picture}
&
\begin{picture}(150,100)
\put(10,10){\vector(2,0){80.}}
\put(10,10){\vector(0,2){80.}}
\put(20,40){\line(3,2){60.}}
\put(80,20){\line(0,2){60.}}
\put(80,40){\oval(120.,40.)[lb]}
\put(0,80){\makebox(0,0){$t$}}
\put(90,0){\makebox(0,0){$\xi$}}
\put(50,-10){\makebox(0,0){\small b)}}
\end{picture}
&
\begin{picture}(100,100)
\put(10,10){\vector(2,0){80.}}
\put(10,10){\vector(0,2){80.}}
\put(10,10){\line(1,2){30.}}
\put(10,10){\line(2,1){60.}}
\put(40,70){\line(2,0){30.}}
\put(70,40){\line(0,2){30.}}
\put(0,80){\makebox(0,0){$t_2$}}
\put(90,0){\makebox(0,0){$t_1$}}
\put(50,-10){\makebox(0,0){\small c)}}
\end{picture}
\end{tabular}
\vspace{0.5cm}
\caption{
}
\label{limint}
\end{figure}

\newpage

\begin{figure}[p]
\unitlength=1mm
\linethickness{0.4pt}
\begin{picture}(156.00,150.00)
\put(9.00,9.00){\framebox(53.00,141.00)[cc]{POLRAD}}
\put(77.00,140.00){\framebox(22.00,10.00)[cc]{INPUT}}
\put(77.00,125.00){\framebox(22.00,10.00)[cc]{CONKIN}}
\put(77.00,110.00){\framebox(22.00,10.00)[cc]{DELTAS}}
\put(77.00,95.00){\framebox(22.00,10.00)[cc]{BORNIN}}
\put(77.00,25.00){\framebox(22.00,65.00)[cc]{QQT}}
\put(77.00,10.00){\framebox(22.00,10.00)[cc]{OUTPUT}}
\put(110.00,95.00){\framebox(21.00,10.00)[cc]{STRF}}
\put(110.00,110.00){\framebox(21.00,25.00)[cc]{F1SFUN}}
\put(110.00,130.00){\makebox(21.00,0)[cc]{G1SFUN}}
\put(110.00,126.00){\makebox(21.00,0)[cc]{G2SFUN}}
\put(110.00,119.00){\makebox(21.00,0)[cc]{F2SFUN}}
\put(110.00,115.00){\makebox(21.00,0)[cc]{B14SF}}
\put(110.00,140.00){\framebox(21.00,10.00)[cc]{}}
\put(110.00,142.00){\makebox(21.00,10.00)[cc]{Models}}
\put(110.00,138.00){\makebox(21.00,10.00)[cc]{for SF}}
\put(110.00,80.00){\framebox(41.00,10.00)[cc]{INTEGRAT}}
\put(110.00,70.00){\framebox(18.00,5.00)[cc]{RV2DL}}
\put(133.00,70.00){\framebox(18.00,5.00)[cc]{RV2}}
\put(122.00,60.00){\framebox(17.00,5.00)[cc]{FFU}}
\put(116.00,50.00){\framebox(29.00,5.00)[cc]{STRF}}
\put(110.00,25.00){\framebox(17.00,5.00)[cc]{AL2LL}}
\put(110.00,35.00){\framebox(17.00,5.00)[cc]{APPTAI}}
\put(127.00,10.00){\framebox(29.00,10.00)[cc]{TARGWS}}
\put(77.00,145.00){\vector(-1,0){15.00}}
\put(62.00,132.00){\vector(1,0){15.00}}
\put(77.00,128.00){\vector(-1,0){15.00}}
\put(99.00,100.00){\vector(1,0){11.00}}
\put(121.00,105.00){\vector(0,1){5.00}}
\put(121.00,135.00){\vector(0,1){5.00}}
\put(99.00,87.00){\vector(1,0){11.00}}
\put(110.00,83.00){\vector(-1,0){11.00}}
\put(119.00,80.00){\vector(0,-1){5.00}}
\put(119.00,75.00){\vector(0,1){5.00}}
\put(142.00,80.00){\vector(0,-1){5.00}}
\put(142.00,75.00){\vector(0,1){5.00}}
\put(136.00,70.00){\vector(0,-1){5.00}}
\put(136.00,65.00){\vector(0,1){5.00}}
\put(119.00,70.00){\vector(0,-1){15.00}}
\put(142.00,70.00){\vector(0,-1){15.00}}
\put(119.00,55.00){\vector(0,1){15.00}}
\put(142.00,55.00){\vector(0,1){15.00}}
\put(125.00,70.00){\vector(0,-1){5.00}}
\put(125.00,65.00){\vector(0,1){5.00}}
\put(99.00,38.00){\vector(1,0){11.00}}
\put(99.00,28.00){\vector(1,0){11.00}}
\put(99.00,45.00){\line(1,0){43.00}}
\put(142.00,45.00){\vector(0,-1){25.00}}
\put(62.00,15.00){\vector(1,0){15.00}}
\put(62.00,74.00){\vector(1,0){15.00}}
\put(77.00,41.00){\vector(-1,0){15.00}}
\put(62.00,117.00){\vector(1,0){15.00}}
\put(77.00,114.00){\vector(-1,0){15.00}}
\put(62.00,102.00){\vector(1,0){15.00}}
\put(77.00,99.00){\vector(-1,0){15.00}}
\end{picture}
\caption{
}
\label{schemPOL}
\end{figure}
\newpage
\begin{figure}[p]
\unitlength=1.00mm
\linethickness{0.4pt}
\begin{picture}(140.00,145.00)
\put(10.00,15.00){\framebox(40.00,130.00)[cc]{SIRAD}}
\put(65.00,135.00){\framebox(30.00,10.00)[cc]{INPUT}}
\put(65.00,120.00){\framebox(30.00,10.00)[cc]{DOS, DOP}}
\put(65.00,105.00){\framebox(30.00,10.00)[cc]{INTEGRAT}}
\put(65.00,90.00){\framebox(30.00,10.00)[cc]{FUS, FUP}}
\put(65.00,75.00){\framebox(30.00,10.00)[cc]{}}
\put(65.00,60.00){\framebox(30.00,10.00)[cc]{INTEGRAT}}
\put(42.00,108.00){\makebox(30.00,10.00)[cc]{\scriptsize
+INTDY}}
\put(42.00,93.00){\makebox(30.00,10.00)[cc]{\scriptsize -INTDY}}
\put(66.00,78.00){\makebox(30.00,10.00)[cc]{FCS, FCP,}}
\put(66.00,73.00){\makebox(30.00,10.00)[cc]{FRS, FRP}}
\put(65.00,45.00){\framebox(30.00,10.00)[cc]{SIGMA}}
\put(65.00,15.00){\framebox(30.00,10.00)[cc]{OUTPUT}}
\put(110.00,90.00){\framebox(30.00,10.00)[cc]{COMVAR}}
\put(110.00,105.00){\framebox(30.00,10.00)[cc]{VPQPK}}
\put(110.00,75.00){\framebox(30.00,10.00)[cc]{DSWS}}
\put(65.00,30.00){\framebox(75.00,10.00)[cc]{}}
\put(104.00,38.00){\makebox(0,0)[cc]{Partonic
distributions}}
\put(104.00,33.00){\makebox(0,0)[cc]{Fragmentation
functions}}
\put(65.00,140.00){\vector(-1,0){15.00}}
\put(50.00,125.00){\vector(1,0){15.00}}
\put(65.00,125.00){\vector(-1,0){15.00}}
\put(50.00,110.00){\vector(1,0){15.00}}
\put(50.00,95.00){\vector(1,0){15.00}}
\put(95.00,95.00){\vector(1,0){15.00}}
\put(80.00,105.00){\vector(0,-1){5.00}}
\put(80.00,90.00){\vector(0,-1){5.00}}
\put(80.00,75.00){\vector(0,-1){5.00}}
\put(80.00,60.00){\vector(0,-1){5.00}}
\put(125.00,100.00){\vector(0,1){5.00}}
\put(125.00,90.00){\vector(0,-1){5.00}}
\put(50.00,20.00){\vector(1,0){15.00}}
\put(104.00,50.00){\vector(0,-1){10.00}}
\put(95.00,50.00){\line(1,0){9.00}}
\end{picture}
\caption{
}
\label{schemSIR}
\end{figure}
\newpage

\begin{figure}[p]
\unitlength=1.00mm
\linethickness{0.4pt}
\begin{picture}(140.00,145.00)
\put(7.00,80.00){\framebox(38.00,65.00)[cc]{AL2LL}}
\put(60.00,135.00){\framebox(37.00,10.00)[cc]{INTEGRAT}}
\put(70.00,120.00){\framebox(27.00,5.00)[cc]{RA2ISS}}
\put(70.00,110.00){\framebox(27.00,5.00)[cc]{RA2IPP}}
\put(70.00,100.00){\framebox(27.00,5.00)[cc]{RA2ISP}}
\put(70.00,90.00){\framebox(27.00,5.00)[cc]{RA2LSS}}
\put(70.00,80.00){\framebox(27.00,5.00)[cc]{RA2LPP}}
\put(70.00,70.00){\framebox(27.00,5.00)[cc]{RA2FSS}}
\put(70.00,60.00){\framebox(27.00,5.00)[cc]{RA2FPP}}
\put(110.00,115.00){\framebox(30.00,10.00)[cc]{SIGMAB}}
\put(110.00,95.00){\framebox(30.00,10.00)[cc]{BOURSC}}
\put(110.00,75.00){\framebox(30.00,10.00)[cc]{Models}}
\put(70.00,40.00){\framebox(27.00,10.00)[cc]{ELUAL2}}
\put(70.00,25.00){\framebox(27.00,10.00)[cc]{ELPAL2}}
\put(110.00,30.00){\framebox(30.00,15.00)[cc]{Form factors}}
\put(45.00,140.00){\vector(1,0){15.00}}
\put(63.00,123.00){\vector(1,0){7.00}}
\put(63.00,113.00){\vector(1,0){7.00}}
\put(63.00,103.00){\vector(1,0){7.00}}
\put(63.00,93.00){\vector(1,0){7.00}}
\put(63.00,83.00){\vector(1,0){7.00}}
\put(63.00,73.00){\vector(1,0){7.00}}
\put(63.00,63.00){\vector(1,0){7.00}}
\put(63.00,45.00){\vector(1,0){7.00}}
\put(63.00,30.00){\vector(1,0){7.00}}
\put(125.00,115.00){\vector(0,-1){10.00}}
\put(125.00,95.00){\vector(0,-1){10.00}}
\put(97.00,45.00){\vector(2,-1){13.00}}
\put(97.00,30.00){\vector(2,1){13.00}}
\put(63.00,135.00){\line(0,-1){72.00}}
\put(104.00,123.00){\line(0,-1){60.00}}
\put(104.00,120.00){\vector(1,0){6.00}}
\put(97.00,123.00){\line(1,0){7.00}}
\put(97.00,113.00){\line(1,0){7.00}}
\put(97.00,103.00){\line(1,0){7.00}}
\put(97.00,93.00){\line(1,0){7.00}}
\put(97.00,83.00){\line(1,0){7.00}}
\put(97.00,73.00){\line(1,0){7.00}}
\put(97.00,63.00){\line(1,0){7.00}}
\put(63.00,63.00){\line(0,-1){33.00}}
\end{picture}
\caption{
}
\label{schemAL2}
\end{figure}
\newpage
\begin{figure}[p]
\unitlength=1.00mm
\linethickness{0.4pt}
\begin{picture}(140.00,145.00)
\put(7.00,80.00){\framebox(38.00,65.00)[cc]{APPTAI}}
\put(60.00,135.00){\framebox(37.00,10.00)[cc]{INTEGRAT}}
\put(110.00,95.00){\framebox(30.00,10.00)[cc]{INTEGRAT}}
\put(110.00,75.00){\framebox(30.00,10.00)[cc]{DIDQG}}
\put(110.00,55.00){\framebox(30.00,10.00)[cc]{Models for SF}}
\put(45.00,140.00){\vector(1,0){15.00}}
\put(125.00,95.00){\vector(0,-1){10.00}}
\put(125.00,75.00){\vector(0,-1){10.00}}
\put(70.00,115.00){\framebox(27.00,10.00)[cc]{}}
\put(83.00,122.00){\makebox(0,0)[cc]{PEAK1}}
\put(83.00,118.00){\makebox(0,0)[cc]{PEAK2}}
\put(110.00,115.00){\framebox(30.00,10.00)[cc]{SIGMAB}}
\put(70.00,95.00){\framebox(27.00,10.00)[cc]{UPRE}}
\put(70.00,65.00){\framebox(27.00,20.00)[cc]{ELP}}
\put(84.00,80.00){\makebox(0,0)[cc]{ELU}}
\put(84.00,70.00){\makebox(0,0)[cc]{ELQ}}
\put(70.00,35.00){\framebox(27.00,20.00)[cc]{Models for SF}}
\put(65.00,75.00){\vector(1,0){5.00}}
\put(65.00,100.00){\vector(1,0){5.00}}
\put(65.00,120.00){\vector(1,0){5.00}}
\put(83.00,65.00){\vector(0,-1){10.00}}
\put(97.00,120.00){\vector(1,0){13.00}}
\put(97.00,100.00){\vector(1,0){13.00}}
\put(65.00,135.00){\line(0,-1){60.00}}
\end{picture}
\caption{
}
\label{schemAPP}
\end{figure}
\newpage
\begin{figure}[p]
\unitlength=1.00mm
\linethickness{0.4pt}
\begin{picture}(115.00,145.00)
\put(10.00,40.00){\framebox(40.00,105.00)[cc]{}}
\put(10.00,40.00){\makebox(40.00,115.00)[cc]{TARGWS}}
\put(10.00,40.00){\makebox(40.00,95.00)[cc]{GWS}}
\put(65.00,135.00){\framebox(35.00,10.00)[cc]{VERCON}}
\put(50.00,140.00){\vector(1,0){15.00}}
\put(65.00,140.00){\vector(-1,0){15.00}}
\put(65.00,120.00){\framebox(35.00,10.00)[cc]{VERTS}}
\put(50.00,125.00){\vector(1,0){15.00}}
\put(65.00,125.00){\vector(-1,0){15.00}}
\put(65.00,105.00){\framebox(35.00,10.00)[cc]{DISTR}}
\put(65.00,90.00){\framebox(35.00,10.00)[cc]{SIGALL}}
\put(65.00,75.00){\framebox(35.00,10.00)[cc]{TTS}}
\put(65.00,60.00){\framebox(35.00,10.00)[cc]{INTEGRAT}}
\put(70.00,40.00){\framebox(26.00,10.00)[cc]{FXI}}
\put(50.00,110.00){\vector(1,0){15.00}}
\put(65.00,110.00){\vector(-1,0){15.00}}
\put(50.00,95.00){\vector(1,0){15.00}}
\put(65.00,95.00){\vector(-1,0){15.00}}
\put(50.00,80.00){\vector(1,0){15.00}}
\put(65.00,80.00){\vector(-1,0){15.00}}
\put(50.00,65.00){\vector(1,0){15.00}}
\put(65.00,65.00){\vector(-1,0){15.00}}
\put(115.00,140.00){\vector(-1,0){15.00}}
\put(115.00,45.00){\vector(-1,0){19.00}}
\put(115.00,125.00){\vector(-1,0){15.00}}
\put(115.00,110.00){\vector(-1,0){15.00}}
\put(82.00,60.00){\vector(0,-1){10.00}}
\put(82.00,50.00){\vector(0,1){10.00}}
\put(115.00,140.00){\line(0,-1){95.00}}
\end{picture}
\caption{
}
\label{schemGWS}
\end{figure}
\newpage

\begin{figure}[p]
\unitlength 1mm
\begin{picture}(160,160)
\put(30,-15){
\epsfxsize=9cm
\epsfysize=12cm
\epsfbox{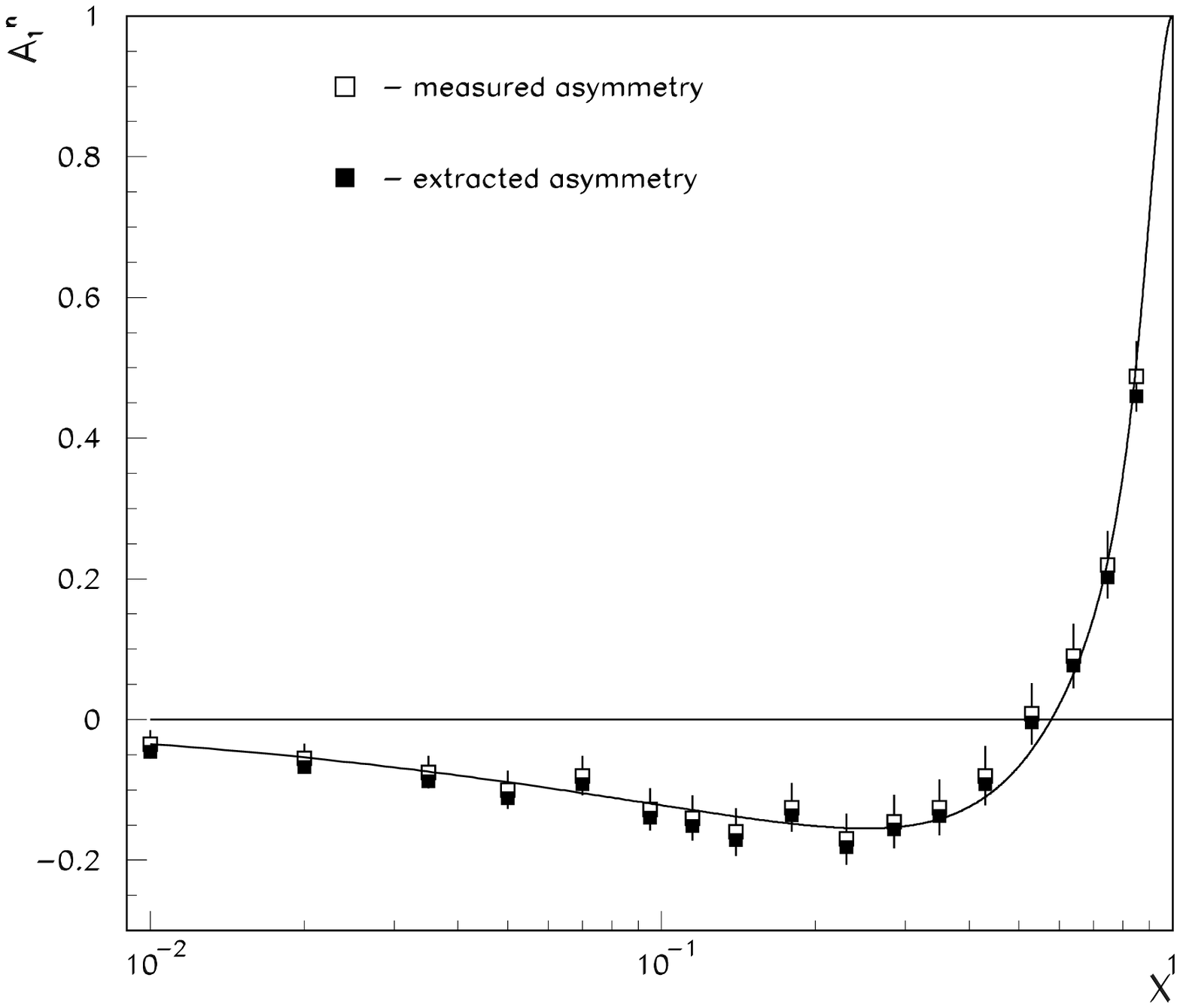}
}
\end{picture}
\caption{
}
\label{fex1}
\end{figure}

\newpage
\begin{figure}[p]
\vspace{4cm}
\unitlength 1mm
\begin{picture}(160,73)
\put(5,-7){
\epsfxsize=7cm
\epsfysize=8cm
\epsfbox{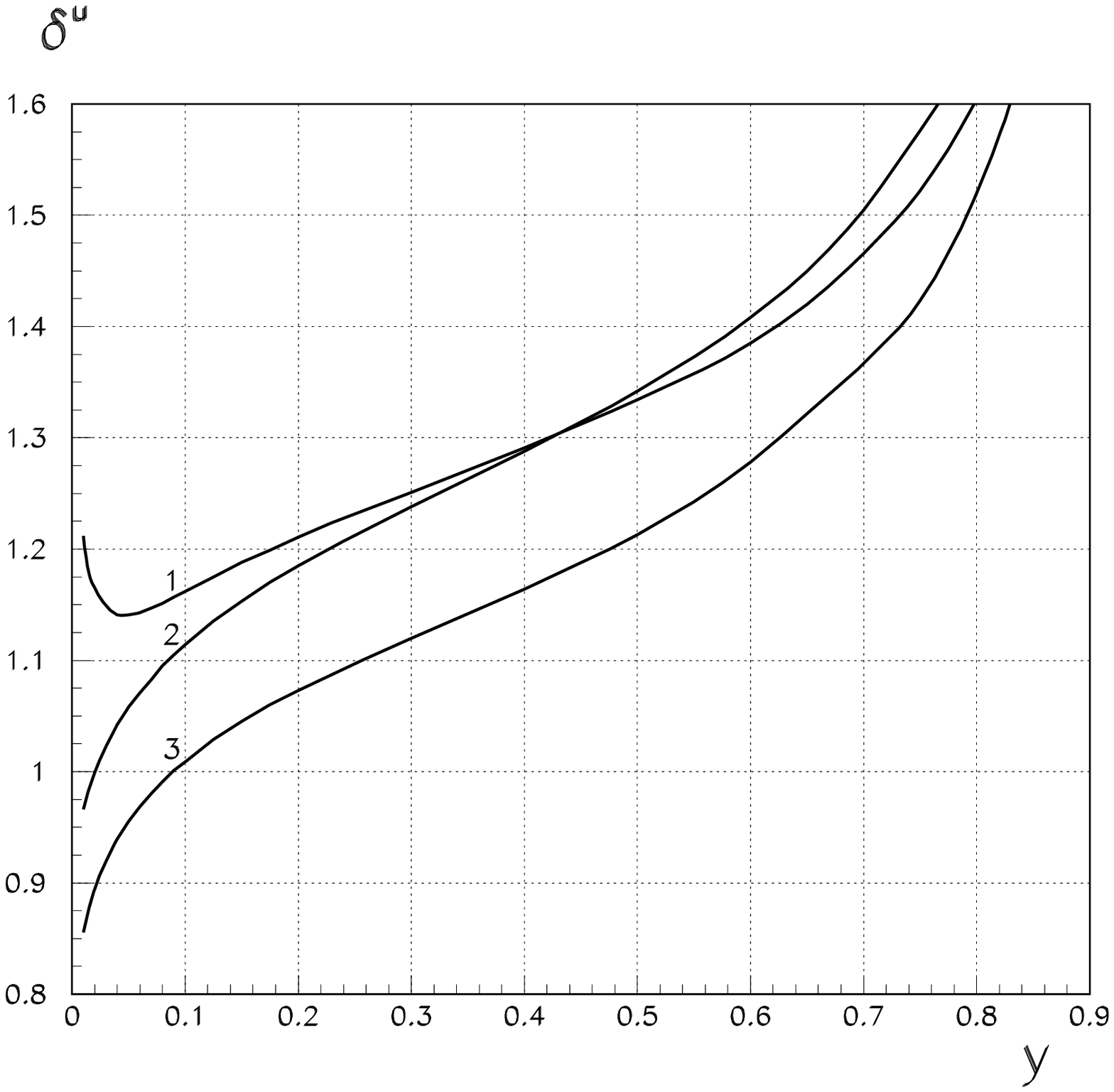}
}

\put(37,0){{\rm a)}}

\put(85,-7){
\epsfxsize=7cm
\epsfysize=8cm
\epsfbox{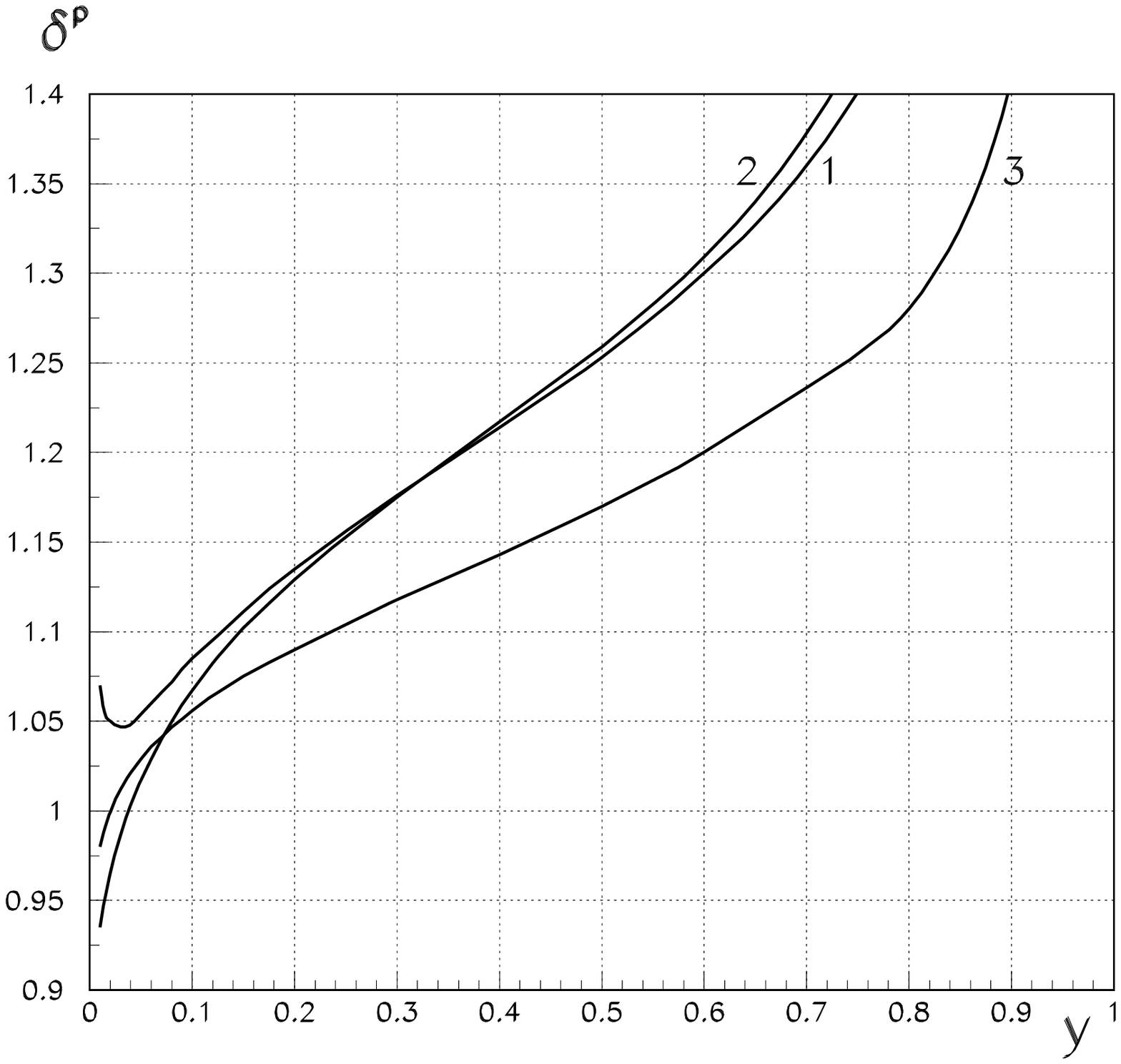}
}
\put(118,0){{\rm b)}}
\end{picture}
\caption{
}
\label{ex2}
\end{figure}
\newpage
\begin{figure}[p]
\vspace{4cm}
\unitlength 1mm
\begin{picture}(160,73)
\put(5,-7){
\epsfxsize=7cm
\epsfysize=8cm
\epsfbox{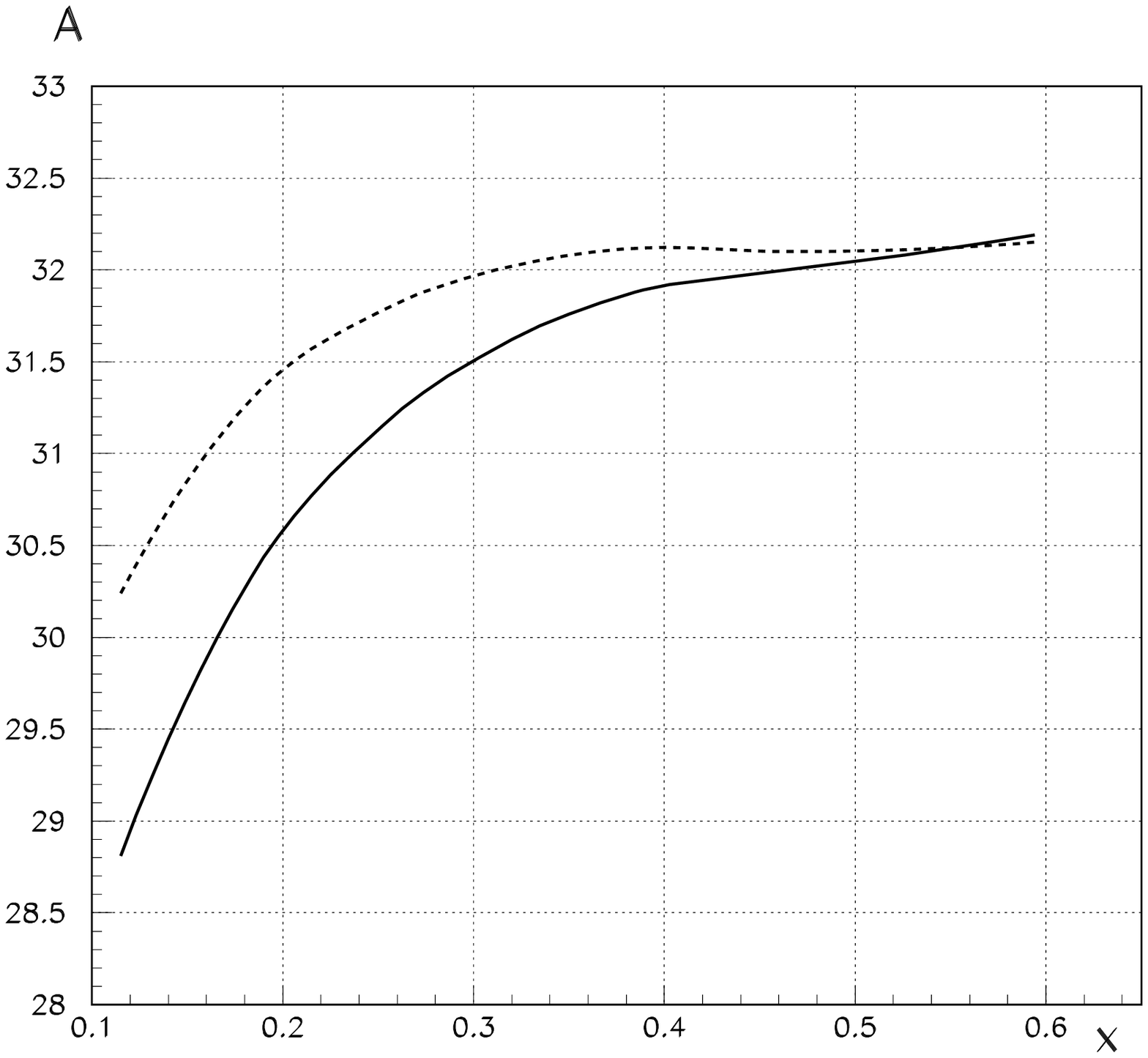}
}

\put(37,0){{\rm a)}}

\put(85,-7){
\epsfxsize=7cm
\epsfysize=8cm
\epsfbox{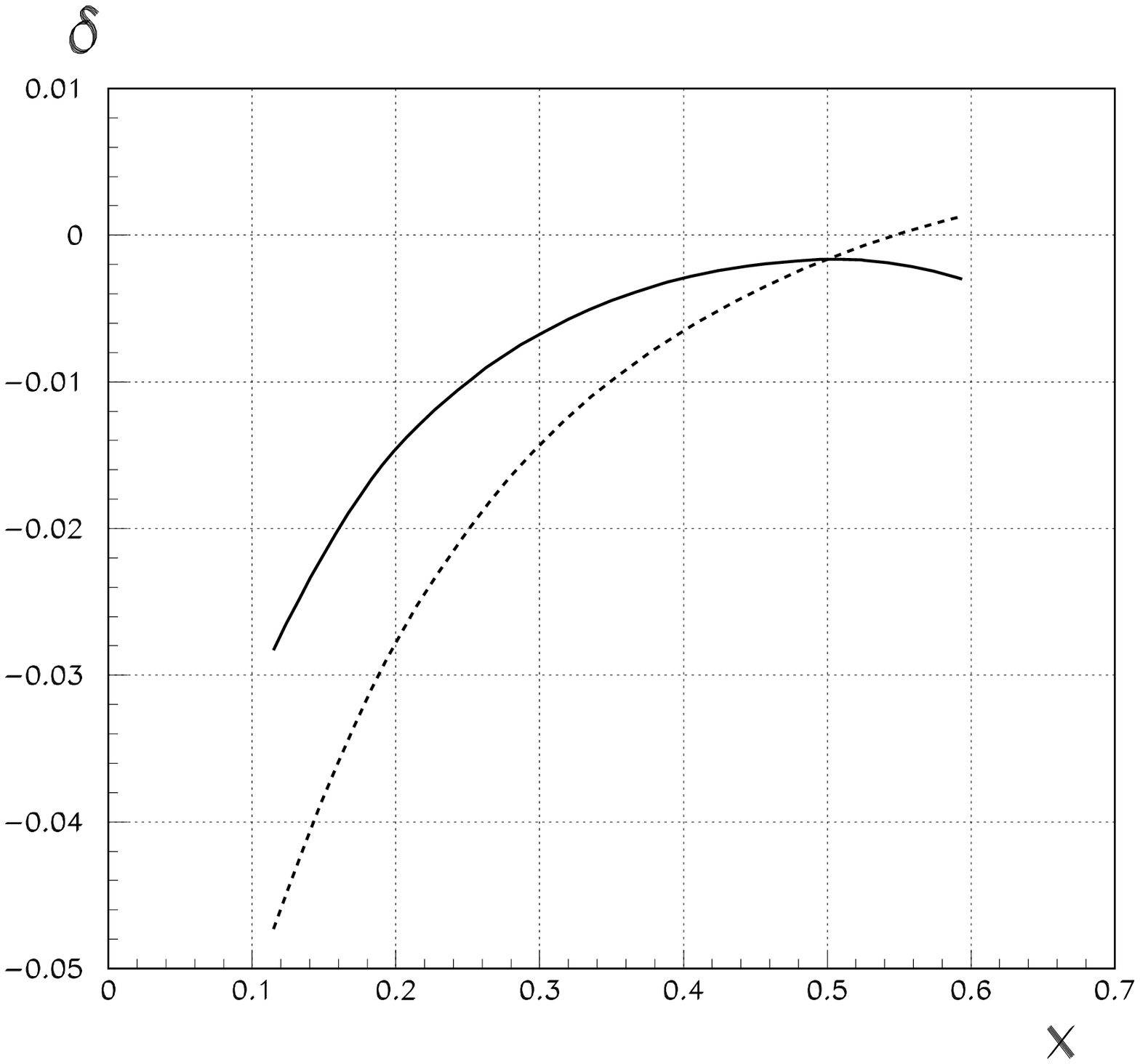}
}
\put(118,0){{\rm b)}}
\end{picture}
\caption{
}
\label{ex3}
\end{figure}
\newpage

\begin{figure}[p]
\vspace{4cm}
\setlength{\unitlength}{0.01in}
\begin{picture}(600,600)(10,-300)
\put(50,50){\vector(0,1){200}}
\put(50,50){\vector(1,0){200}}
\put(50,100){\line(1,0){200}}
\put(150,50){\line(0,1){50}}
\put(20,95){0.3}
\put(140,30){1.6}
\put(35,30){0}
\put(130,260){$R(Q^2,W^2)$}
\put(20,250){$Q^2$,}
\put(10,230){$GeV^2$}
\put(210,30){$W^2,\;GeV^2$}
\put(170,170){\cite{Whi}}
\put(60 ,70 ){\cite{Bra}-$R=0$}
\put(160,70 ){\cite{Ste}-$R=0.18$}
\put(60 ,110){\line(1,1){25}}
\put(60 ,130){\line(1,1){25}}
\put(60 ,150){\line(1,1){25}}
\put(60 ,170){\line(1,1){25}}
\put(60 ,190){\line(1,1){25}}
\put(60 ,210){\line(1,1){25}}
\put(60 ,230){\line(1,1){25}}
\put(350,50){\vector(0,1){200}}
\put(350,50){\vector(1,0){200}}
\put(350,140){\line(1,0){100}}
\put(450,100){\line(1,0){100}}
\put(450,50){\line(0,1){90}}
\put(330,135){6}
\put(445,30){4}
\put(335,30){0}
\put(430,260){$F_2^p(Q^2,W^2)$}
\put(320,250){$Q^2$,}
\put(310,230){$GeV^2$}
\put(470,30){$W^2,\;GeV^2$}
\put(470,170){\cite{Mil}}
\put(395 ,90 ){\cite{Bra}}
\put(465,70 ){\cite{Ste}}
\put(360 ,150){\line(1,1){25}}
\put(360 ,170){\line(1,1){25}}
\put(360 ,190){\line(1,1){25}}
\put(360 ,210){\line(1,1){25}}
\put(360 ,230){\line(1,1){25}}
\put(535,55 ){\line(1,1){15}}
\put(535,65 ){\line(1,1){15}}
\put(535,75 ){\line(1,1){15}}
\put(535,85 ){\line(1,1){15}}

\put(50,-250){\vector(0,1){200}}
\put(50,-250){\vector(1,0){200}}
\put(50,-100){\line(1,0){100}}
\put(150,-250){\line(0,1){150}}
\put(150,-200){\line(1,0){100}}
\put(20,-105){30}
\put(145,-270){7}
\put(35,-270){0}
\put(110,-40){${F_2^n(Q^2,W^2)\over F_2^p(Q^2,W^2)}$}
\put(20,-50){$Q^2$,}
\put(10,-70){$GeV^2$}
\put(210,-270){$W^2,\;GeV^2$}
\put(190,-110){\cite{Mil}}
\put(100 ,-175){\cite{Whi}}
\put(180,-230 ){\cite{Ste}}
\put(60 ,-240){\line(1,1){25}}
\put(60 ,-220){\line(1,1){25}}
\put(60 ,-200){\line(1,1){25}}
\put(60 ,-180){\line(1,1){25}}
\put(60 ,-160){\line(1,1){25}}
\put(60 ,-140){\line(1,1){25}}
\put(60 ,-120){\line(1,1){18}}
\put(60 ,-95){\line(1,1){25}}
\put(60 ,-75){\line(1,1){25}}
\put(60 ,-55){\line(1,1){25}}
\put(235,-245 ){\line(1,1){15}}
\put(235,-235 ){\line(1,1){15}}
\put(235,-225 ){\line(1,1){15}}
\put(235,-215 ){\line(1,1){15}}
\end{picture}
\caption{
}
\label{SFF2}
\end{figure}

\end{document}